\documentclass[11pt]{article}

\usepackage{epsfig}
\usepackage{graphicx}
\usepackage{latexsym}
\usepackage{amsmath}
\usepackage{amsbsy}
\usepackage{amssymb}
\usepackage{theorem}
\usepackage{amsfonts}

\usepackage[francais,english]{babel}
\usepackage[latin1]{inputenc}
\usepackage[T1]{fontenc}

\def\be{\begin{equation} \displaystyle}
\def\ee{\end{equation} }

\def\bea{\begin{eqnarray}}
\def\eea{\end{eqnarray} }
\def\bean{\begin{eqnarray*}}
\def\eean{\end{eqnarray*} }

\def\div{{\rm div }}

\def\R{{\rm I\hspace{-0.50ex}R} }
\def\C{\rm \hbox{C\kern-.57em\raise.47ex
         \hbox{$\scriptscriptstyle |$}\kern+0.5 em }}
\def\E{{\rm I\hspace{-0.50ex}E} } %esp\'erance
\def\P{{\rm I\hspace{-0.50ex}P} } %probabilit\'e, \'el\'ements finis
\def\tr{\mathrm{tr}\,} %trace d'une matrice
\def\Var{\mathrm{Var}} %variance d'une variable al\'eatoire
\def\I{\text{Id}} % l'identité
\def\ttau{\tilde{\tau}}

\def\ou{\overline{u}}
\def\oP{\overline{P}}
\def\oQ{\overline{Q}}
\def\otau{\overline{\tau}}

\def\opsi{\overline{\psi}}

\def\SS{\bold{S}}

\def\ttau{\boldsymbol{\tau}}
\def\ssigma{\boldsymbol{\sigma}}

\def\osc{\rm osc}

\newtheorem{theo}{Theorem}
\newtheorem{prop}{Proposition}

\newtheorem{rem}{Remark}

\newcommand{\uu}[1]  {{\boldsymbol #1} }
\newcommand{\dps}{\displaystyle   }

\def\FF{\uu{F}}
\def\XX{\uu{X}}

\def\tXX{\tilde{\XX}}
\def\WW{\uu{W}}
\def\BB{\uu{B}}
\def\We{\text{We}}
\def\Rey{\text{Re}}

\def\pp{\uu{p}}
\def\qq{\uu{q}}
\def\PP{\uu{P}}
\def\QQ{\uu{Q}}
\def\xx{\uu{x}}
\def\RR{\uu{R}}

\newcommand{\U}      { \uu{u}         }
 
\newcommand{\V}      { \uu{v}         }
\newcommand{\jump}[1]{[\![#1]\!]}

\title{Micro-macro models for viscoelastic fluids: modelling, mathematics and numerics}
\author{C. Le Bris, T. Lelièvre\\
\footnotesize{ CERMICS, Ecole Nationale des
  Ponts (ParisTech), 6 \& 8 Av. B. Pascal,
  77455 Marne-la-Vall\'ee, France.}\\
\footnotesize{ INRIA Rocquencourt, MICMAC team, B.P. 105, 78153 Le Chesnay Cedex, France.}\\
 lebris@cermics.enpc.fr, lelievre@cermics.enpc.fr
}

\begin{document}

\maketitle

\abstract{This paper is an introduction to the modelling of viscoelastic fluids, with an emphasis on micro-macro (or multiscale) models. Some elements of mathematical and numerical analysis are provided. These notes closely follow the lectures delivered by the second author at the Chinese Academy of Science during the Workshop "Stress Tensor Effects on Fluid Mechanics", in January 2010.}

\tableofcontents

\section{Introduction}

Many fluids exhibit a behaviour very different from that of air or water, in particular because of some memory effects. Such fluids are called {\em complex} or {\em non-Newtonian} fluids. Examples include blood, muds, fresh concrete, paints, etc... In many cases, this peculiar behaviour originates from the presence of some microstructures within the fluid the evolutions of which are strongly coupled to the solvent dynamics. In the following, we focus on dilute solutions of flexible polymer chains. This is a prototypical example of such fluids, both from an experimental and a modelling viewpoint. A dilute solution of polymer chains consists of a solvent and polymer chains floating therein. These chains are in such a small quantity that direct interactions between the chains can be neglected. In the following, a polymer chain is meant to be a long linear molecule built as the repetition of an elementary pattern, called the monomer (think typically of an alcane molecule $CH_3-(CH_2)^n-CH_3$). It is observed that the rheology of the fluid (that is the way it flows) is very much affected by the polymer chains, even at a very small concentration. This paper is an introduction to microscopic-macroscopic (in short micro-macro) models to describe such fluids. The modelling principle is to couple conservation laws on macroscopic quantities (such as the velocity or the stress) with some models for the evolutions of microstructures. Direct macroscopic approaches will be also discussed.

We assume that the reader is familiar with standard analytical techniques and numerical methods for partial differential equations. Our previous work~\cite{le-bris-lelievre-09} can be seen as a companion paper, more intended to students less knowledgeable on numerical and mathematical analysis for partial differential equations. We would also like to mention the more general overview~\cite{blanc-legoll-le-bris-lelievre-10} on multiscale methods (not only for polymeric fluids, but also for solid materials for example).

The paper is organized as follows. In Section~\ref{sec:model}, we present and discuss the building blocks of a micro-macro model for a complex fluid, with an emphasis on polymeric fluids. This Section contains in particular an introduction to the kinetic modelling of polymer chains, and to coarse-graining techniques based on the notion of free energy. Section~\ref{sec:math} is an overview of the mathematical analysis results that have been obtained on models for viscoelastic fluids in general, and micro-macro models in particular. In that section, an introduction to entropy techniques for longtime behaviour of partial differential equations is provided. Finally, Section~\ref{sec:num} is devoted to various problems related to the numerical discretization of models for viscoelastic fluids: free energy dissipative schemes for macroscopic models, an introduction to a nonlinear approximation approach to solve high-dimensional partial differential equations, and various techniques for reducing the variance of the numerical results.

\section{Modelling}\label{sec:model}

The focus of this work is complex fluids whose non-Newtonian behaviour is due to some microstructures within the fluid. One example is blood: red blood cells may aggregate and form complex structures that affect the rheology of the blood.

More precisely, we are interested in cases where:
\begin{enumerate}
\item the microstructures are very numerous (per unit volume, say),
\item the microstructures are small and light,
\item the solvent is Newtonian.
\end{enumerate}
The first characteristic will be important in the following to express macroscopic quantities as means over many configurations of the microstructures, using statistical mechanics techniques. The second characteristic will be useful to describe the evolution of the microstructures using the Langevin equations, thermalization being modeled using a Brownian motion in the evolution equations. Finally, the third characteristic is required in order to write the equations at the macroscopic level, assuming that the only non-Newtonian contribution to the fluid originates from the microstructures. These assumptions do not cover all non-Newtonian fluids with micro-structures, a typical counter-example being granular materials, for which other approaches are required. However, such a setting is useful for a wide range of materials, a prototypical example being dilute solution of polymer chains, which is the main example considered in this paper. This setting also provides a convenient test-bed for arguments in mathematical analysis and approaches for numerical simulations that may prove useful elsewhere.

\subsection{Experimental observations}

Non-Newtonian fluids are ubiquitous: food industry (mayonnaise, egg white, jellies)~; materials industry (plastic or polymeric fluids)~; biology or medicine (blood, synovial liquid)~; civil engineering (fresh concrete, paints)~; environment (snow, muds, lava)~; cosmetics (shaving cream, toothpaste, nail polish)~; etc... Non-Newtonian fluids may exhibit very weird behaviours, such as, for polymeric fluids, the open syphon effect, or the rod climbing effect (also called the Weissenberg effect).

\begin{figure}
\centerline{\includegraphics[height=4cm]{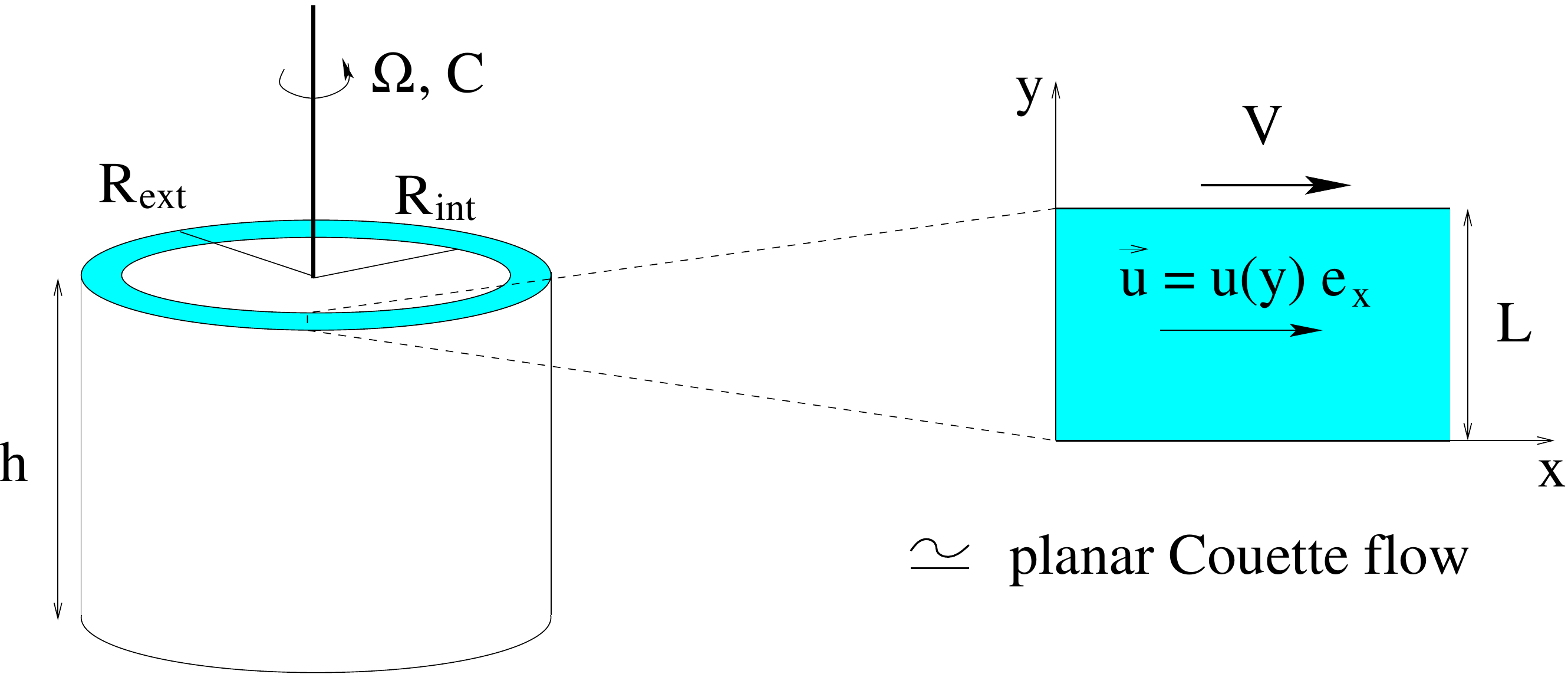}}
\caption{Schematic representation of a rheometer. The shear stress is given by $\tau=\frac{C}{2 \pi R_{int}^2 h}$ where $C$ is the torque and the shear rate is $\dot{\gamma}=\frac{V}{L}=\frac{R_{int} \Omega}{R_{ext}- R_{int}}$ where $\Omega$ is the angular velocity.}\label{fig:shear}
\end{figure}

The experimental approach to study Non-Newtonian fluids consists in measuring the response of the fluid (the stress) under  specific constraints (typically homogeneous flows), in order to characterize the stress-strain rate relation. A typical experimental setting is a simple shear flow obtained in appropriate rheometers, see Figure~\ref{fig:shear}. The stress-strain rate relation is obtained by measuring the torque as a function of the angular velocity, at a stationary state. The shear stress $\tau$ is measured for various values of the shear rate $\dot{\gamma}$. One can thus discriminate between four typical behaviours: Newtonian fluids, for which $\tau=\eta \dot{\gamma}$, $\eta$ being the shear viscosity ; Shear-thinning fluids, for which the effective viscosity $\tau/\dot{\gamma}$ diminishes as $\dot{\gamma}$ increases ; Shear-thickening fluids, for which the effective viscosity $\tau/\dot{\gamma}$ increases as $\dot{\gamma}$ decreases ; Yield-stress fluid, for which $\dot{\gamma}$ remains $0$ until a threshold on the stress $\tau$. As explained below, the consideration of such simple flows allows to formally classify the non-Newtonian fluids, and to propose values for some parameters appearing in macroscopic laws. Of course, the challenge is then to assess the validity of these macroscopic laws for more complex flows.

\subsection{Multiscale modelling}

Modelling of non-Newtonian fluids starts with the mass and momentum conservation equations for incompressible fluids:
\begin{equation}\label{eq:momentum}
\left\{
\begin{aligned}
 \rho \left( \frac{\partial \U}{\partial t}  + \U.\nabla \U \right) &= - \nabla p + \div
 (\ssigma),\\
 \div(\U)&=0,
\end{aligned}
\right.
\end{equation}
supplied with appropriate boundary and initial conditions.
Throughout this article, $\U$ is the velocity, $p$ the pressure, $\rho$ is the density of the fluid and $\ssigma$ is the stress. For Newtonian fluids, one postulates a linear relation between the stress and the strain rate:
\begin{equation*}
\ssigma=\eta(\nabla \U + \nabla \U^T),
\end{equation*}
$\eta$ being the viscosity of the fluid.
For non-Newtonian fluids, an additional term (the so-called {\em extra-stress}) is introduced in the constitutive relation
\begin{equation}\label{eq:non-Newt}
\ssigma=\eta(\nabla \U + \nabla \U^T) + \ttau.
\end{equation}
The question is now how to define an evolution law on $\ttau$, as a function of the other macroscopic quantities.

\begin{figure}
\centerline{\includegraphics[height=6cm]{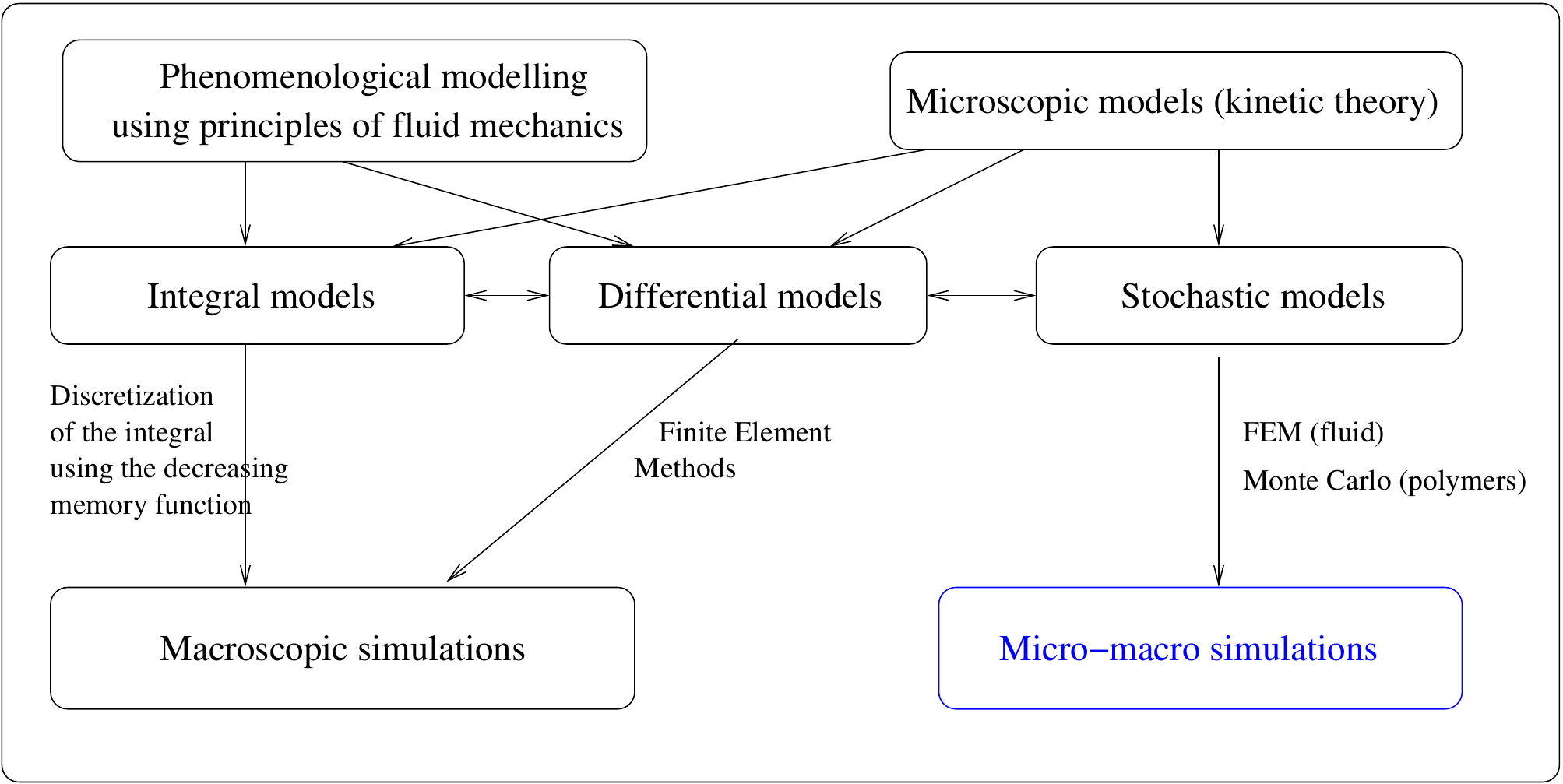}}
\caption{Classical approaches for the modelling of complex fluids.}\label{fig:model}
\end{figure}

There are basically two approaches for the derivation of such a relation, see Figure~\ref{fig:model}. The first one is to postulate an equation relating $\ttau$ and $\nabla \U$ (using typically frame invariance as a guideline) and to fit the parameters to simple experiments, as explained above. This leads to two classes of models: (i) {\em integrals models}, which provide the stress at a given time and a given point as an integral over the past of the trajectory which leads to this point:
$$\ttau=\int_{-\infty}^t m(t-t') \SS_t(t') \, dl(t')$$
where $\SS_t(t')$ is some tensor defined as a function of $\U$ (ii) {\em differential models}, which are partial differential equations involving~$\U$ and~$\ttau$, a prototypical example being the Oldroyd-B model (that we will discuss in more details later on):
\begin{equation} \label{eq:OB}
  \lambda \left(\frac{\partial \ttau}{\partial t} + \U \cdot \nabla \ttau - \nabla \U \ttau - \ttau \nabla \U^T \right) + \ttau = \eta_p (\nabla \U + \nabla \U^T),
\end{equation}
where $\lambda$ and $\eta_p$ are two parameters. This first approach thus yields so-called macroscopic models. They only involve macroscopic quantities (the velocity, the pressure and the stress). 

Another route (that we will follow below) is to model the behaviour of the microstructures responsible for the non-Newtonian features of the fluid, and to obtain the stress $\ttau$ as a function of the configurations of these  microstructures. In such an approach, additional variables are required (to model the state of the microstructures), and one thus obtains higher dimensional systems to simulate. The complete model thus couples the Equations~\eqref{eq:momentum} (for mass and momentum conservations on macroscopic quantities) to some equations describing the evolution of the microscopic structures. This forms a so-called {\em micro-macro (or multiscale) approach}. This approach may be seen as a way to justify (or find new) macroscopic models, giving a precise substance to the parameters that appear in those models. It may also be seen as a modelling tool in its own rights, when the micro-macro model does not lead to an equivalent macroscopic model, and is discretized as such.

Multiscale approaches are now very popular for the modelling of materials (fluids or solids). Such techniques raise several questions, from the modelling viewpoint, numerically and theoretically. Theoretically, the main question is the consistency of the approach, namely the fact that the models used at various levels are compatible. More precisely, one would  typically like the microscopic models to be refinements upon purely macroscopic models. This raises the question of deriving macroscopic models from microscopic models (typically using closure approximations, homogenization, etc...). Numerically, many questions are related to the efficiency of the coupling procedures ({\em a posteriori} error estimators and adaptive refinement techniques to use the finest model only when needed, adequate methods to couple models different in nature, such as deterministic and stochastic models, etc...).

From a modelling viewpoint, the delicate question is the discrepancy of the time and length scales. Typically, the macroscopic and the microscopic descriptions do not share the same time and length scales. The finest time and length scales (at the microscopic level) should not be resolved, otherwise the macroscales cannot be reached. Two techniques are used. For timescales, one may use {\em local thermodynamic equilibrium assumptions} to consider that the microscopic configurations reach a stationary state (given the macroscopic quantities) within a timescale much smaller than the macroscopic timescale. For lengthscales, one may use {\em mean field approximations}, which consist in replacing the many-particle system at the microscopic level by a one-particle problem within an averaged environment. Under these approximations, the macroscopic quantities are typically obtained as averages over microscopic quantities (locally at the macroscopic time and length scale). Notice that for both time and space, these approximations are based on scales separation arguments. Besides, from a computational viewpoint, {\em coarse-graining techniques} are useful in such approaches to reduce the number of degrees of freedom at the finest scale. Typically, coarse-graining implies an increase of the microscopic characteristic time and length scales. All these techniques will be illustrated below on the modelling of polymeric fluids.

In the simplest situations, the outcome of the above modelling is a one-way coupled model (information passing): the microscopic model is simply used to feed the macroscopic model, by giving values to some parameters (for example, the viscosity of a fluid in the Navier-Stokes equations may be obtained using molecular dynamics). But there also exist more complicated situations, where the mean field approximation to replace the many-body system by a one-body problem at the microscopic level is not possible (this is the case of granular materials for example), or where the local thermodynamic equilibrium assumption is not valid. In the latter situation, it is not possible to assume that the microstructures completely relax to a local equilibrium (the micro time scale being of the same order as the macro timescale), and then one obtains genuinely coupled systems (information coupling), as is the case for polymeric fluids.

\subsection{Microscopic models for polymer chains}

Let us momentarily assume that the velocity field $\U$ of the solvent is fixed and given. The question we address in this section is the modelling of the polymer chain dynamics. As explained above, we need a description both sufficiently fine to get the essential features of a polymer chain that are of interest to get the correct macroscopic behaviour (the correct rheology of the fluid) and sufficiently coarse so that the microscopic model is not computationally prohibitively expensive. In the following, we only consider a linear chain (without branching). Classical references for the modelling procedure described below are~\cite{doi-edwards-88,oettinger-95,BCAH-87-1,BCAH-87-2}.

\subsubsection{Modelling of a chain in the solvent at rest}

We first assume that $\U=0$. A standard technique to model the evolution of a polymer chain is based on molecular dynamics. It amounts to assuming that the interactions between the nuclei (these interactions would ideally be obtained using electronic structure computations) can be modelled by a potential function, and then to writing a dynamics (say, Newton equations of motion) on the nuclei. Let us assume that the polymer chain contains $N$ atoms, with positions $\uu{q}=(\uu{q}_1,\dots,\uu{q}_N) \in \R^{3N}$, interacting through a potential function $V:\uu{q} \in \R^{3N} \mapsto V(\uu{q}) \in \R$. In order to avoid expensive electronic structure computations, empirical models of the form
$$V(\uu{q}_1,...,\uu{q}_N)=\sum_{i<j} V_{\rm pair}(\uu{q}_i,\uu{q}_j)
+ \sum_{i<j<k} V_{\rm triplet}(\uu{q}_i,\uu{q}_j,\uu{q}_k) + \ldots$$
are used for the potential $V$. For a polymer chain, one would typically have to consider pair interactions related to covalent bonds, and interactions involving three or four consecutive Carbon atoms along the backbone of the polymer since the potential energy depends on angles and dihedral angles along the chain. A standard molecular dynamics is then given by the Langevin equations:
\begin{equation}\label{eq:lang}
\left\{
\begin{array}{l}
d \QQ_t = M^{-1} \PP_t \, dt, \\
d \PP_t = -\nabla V(\QQ_t)\, dt   - \zeta  M^{-1} \PP_t \, dt + \sqrt{2 \zeta \beta^{-1}} d\WW_t ,
\end{array}
\right.
\end{equation}
where $\PP_t$ is the momentum, $M$ is the mass tensor, $\zeta$ is a friction coefficient and $\beta^{-1}= k T$. Here, $\WW_t$ denotes a $3N$-dimensional Brownian motion. When $\zeta=0$, these equations simplify into the Newton equations (Hamiltonian system). The two terms depending on~$\zeta$ model the effect of the bath, namely the thermalization due to the many-collisions of the polymer chain with the surrounding solvent molecules. The term $- \zeta  M^{-1} \PP_t \, dt$ models a (viscous) friction and the term $\sqrt{2 \zeta \beta^{-1}} d\WW_t$ models thermal agitation. These two terms balance each other (the friction rate and the diffusion constant are related through {\em the fluctuation dissipation relation}) so that the canonical Boltzmann-Gibbs probability measure:
$$\nu(d \pp,d \qq)=\overline{Z}^{-1} \exp\left(-\beta \left(\frac{\pp^T M^{-1} \pp}{2} + V (\qq)  \right) \right) \, d\pp \, d\qq$$
is left invariant by the dynamics. Notice that $\nu(d \pp,d \qq)$ is the product of a Gaussian measure on the momentum $\pp$ with the probability measure
$$\mu(d\qq)=Z^{-1} \exp(-\beta V(\qq)) \, d\qq$$
 on the positions. Below, we will need the so-called {\em zero mass Langevin dynamics}, which is the approximation
$$0= -\nabla V(\QQ_t)\, dt  - \zeta d \QQ_t + \sqrt{2 \zeta \beta^{-1}} d\WW_t$$
 of~\eqref{eq:lang} obtained by letting the mass go to zero. The zero-mass Langevin equation is thus a stochastic differential equation on the positions only:
\begin{equation}\label{eq:0mlang}
d\QQ_t= -\zeta^{-1} \, \nabla V(\QQ_t)\, dt + \sqrt{2 \zeta^{-1} \beta^{-1}} d\WW_t.
\end{equation}
This dynamics leaves the canonical measure $\mu$ invariant.

This fine description of a polymer chain is not suitable for a multiscale model, since it is too high-dimensional. For computational purposes, it is thus desirable to derive a coarse-grained model. The assumption underlying the simplest such model, the so-called {\em dumbbell model}, is that essentially the orientation and the length of the polymer chain suffice to describe the influence of the microstructures on the rheology of the complex fluid. Let us thus introduce the so-called end-to-end vector
$$
\xi:
\left\{
\begin{array}{ccc}
\R^{3N} & \to & \R^3\\
\qq=(\qq_1, \ldots, \qq_N) & \mapsto & \xx=\qq_N - \qq_1
\end{array}
\right.
$$
The function $\xi$ is called a {\em collective variable}. To obtain an effective potential, a classical procedure is to introduce the so-called {\em free energy} $\Pi(\xx)$ which is such that
\begin{equation}\label{eq:FE_def}
\xi * \left( Z^{-1}\exp(-\beta V(\qq)) \, d\qq \right) = \exp(-\beta \Pi(\xx)) \, d\xx,
\end{equation}
where $\xi*\left( Z^{-1}\exp(-\beta V(\qq)) \, d\qq \right)$ is the image of the canonical measure $\mu$ by $\xi$. An explicit definition is
$$\Pi(\xx) = -\beta^{-1} \ln \left( \int \exp(-\beta V(\qq)) \delta_{\xi(\qq)-\xx}(d\qq) \right),$$
where the measure $\delta_{\xi(\qq)-\xx}(d\qq)$ (with support $\{ \qq, \xi(\qq)=\xx\}$) is defined by the conditioning formula $d\qq=\delta_{\xi(\qq)-\xx}(d\qq) \, d\xx$. The free energy $\Pi$ is thus built so that the equilibrium thermodynamic properties are exactly conserved, namely, for any (bounded) function $\varphi : \R^{3} \to \R$
\begin{equation}\label{eq:FE_def_}
\int \varphi( \xi(\qq)) \mu(d\qq) = \int \varphi(\xx) \exp(-\beta \Pi(\xx)) \, d\xx,
\end{equation}
a property which is equivalent to~\eqref{eq:FE_def}. For more details on free energy, we refer to~\cite{lelievre-rousset-stoltz-book-10}.

For polymeric fluids, the definition of the potential $V$ to model the polymer chain is based upon the Kramers model which describes a freely jointed bead-rod chain. Using the coarse-graining procedure described above, one obtains a potential $\Pi$ (the so-called inverse Langevin potential) which is in practice approximated by:
\begin{itemize}
\item either the {\em FENE potential}:
$$\Pi_{\rm FENE}(\xx) = \frac{-b }{2\beta} \ln \left( 1- \frac{\beta H |\xx|^2}{b} \right),$$
\item or the {\em Hookean potential}:
$$\Pi_{\rm HOOK}(\xx) = \frac{\beta H |\xx|^2}{2}.$$
\end{itemize}
The parameters $H$ and $b$ depend on the number of segments in the Kramers chain. The acronym FENE stands for {\em Finitely Extensible Nonlinear Elastic}: the FENE model takes into account the finite extensibility of the chain, while the Hookean model is only a correct approximation of the inverse Langevin potential around $|\xx|=0$.

\begin{figure}
\centerline{\includegraphics[height=4cm]{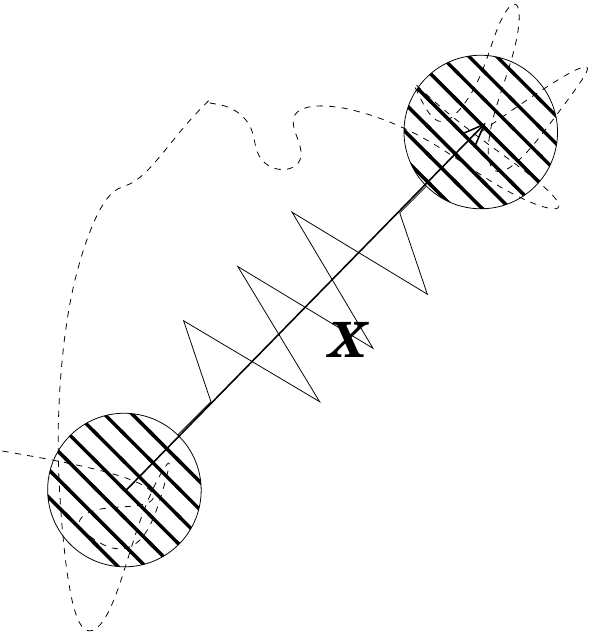}}
\caption{In the dumbbell model, the polymer chain is modelled by two beads linked by a spring.}\label{fig:Atom2dumb}
\end{figure}

The dumbbell model thus consists in modelling the chain by two beads linked by a spring with potential $\Pi$, see Figure~\ref{fig:Atom2dumb}.
Notice that more elaborate coarse-grained models have also been introduced, using more than one vector to more precisely describe the shape of the polymer chain by a chain of beads linked by springs. The derivation we describe here and below apply to such models.

\subsubsection{Modelling of a chain in the moving solvent}

We now assume that the velocity $\U$ of the surrounding solvent is still given but non-zero. The evolution of the dumbbell (the coarse-grained model of the polymer chain) consists in writing the zero-mass Langevin on the two beads of the dumbbell (that is, the two ends of the chain). In the following, we use the notation $\XX^1_t$ and $\XX^2_t$ for the positions of the two beads (instead of $\QQ_{1,t}$ and $\QQ_{N,t}$, which would be more consistent with the latter section) since the beads are thought to be the centers of mass of a few monomers, at each end of the polymer chain, rather than the very ends of the chain. Then, for $i=1,2$, one has
$$
\left\{
\begin{aligned}
&d\XX^1_t- \U(t,\XX^1_t) \, dt = - \zeta^{-1} \FF(\XX^1_t-\XX^2_t) \, dt + \sqrt{2 \beta^{-1} \zeta^{-1}} d \BB^1_t, \\
&d\XX^2_t- \U(t,\XX^2_t) \, dt = - \zeta^{-1} \FF(\XX^2_t-\XX^1_t) \, dt + \sqrt{2 \beta^{-1} \zeta^{-1}} d \BB^2_t,
\end{aligned}
\right.
$$
where $\BB^1_t$ and $\BB^2_t$ are two independent $3$-dimensional Brownian motions.
This is the zero-mass Langevin equations~\eqref{eq:0mlang} written on each bead, using the fact that the friction force is now $-\zeta( d\XX^i_t - u(t, \XX^i_t) \, dt)$ since the solvent is moving with velocity $\U$, and the interaction potential is $\Pi(\xx_1-\xx_2)$. In these equations, we have introduced the notation
$$\FF(\xx)=\nabla \Pi(\xx)$$
for the so-called {\em entropic force} (or {\em mean force}) which is the derivative of the free energy $\Pi$.  Let us introduce the end-to-end vector
$$\XX_t=\XX^2_t-\XX^1_t$$
and the position of the center of mass
$$\RR_t=\frac{\XX^1_t+\XX^2_t}{2}.$$
By linear combination, we obtain
\begin{equation*}
\left\lbrace
\begin{aligned}
&\displaystyle {d \XX_t = \left( \U(t,\uu{X}^2_t) - \U(t,\uu{X}^1_t) \right) \, dt - 2\zeta^{-1}
\FF(\XX_t) \, dt + 2 \sqrt{\beta^{-1}\zeta^{-1}} d \WW^1_t},\\
&\displaystyle {d \uu{R}_t = \frac{1}{2}\left( \U(t,\uu{X}^1_t) + \U(t,\uu{X}^2_t) \right) \, dt + \sqrt{\beta^{-1}\zeta^{-1}} d \WW^2_t},
\end{aligned}
\right.
\end{equation*}
where  $\uu{W}^1_t=\frac{1}{\sqrt{2}}\left( \uu{B}^2_t - \uu{B}^1_t\right)$
and $\uu{W}^2_t=\frac{1}{\sqrt{2}}\left( \uu{B}^1_t + \uu{B}^2_t\right)$ are two independent Brownian motions. We now use two approximations:
 \begin{itemize}
 \item an expansion of the velocity field: 
\begin{equation}\label{eq:exp_u}
\U(t,\XX^i_t) \simeq \U(t,\RR_t) + \nabla \U(t,\RR_t)
   (\XX^i_t - \RR_t),
\end{equation}
 \item a smallness assumption on the noise on $\RR_t$:  \begin{equation}\label{eq:small_noise}
\sqrt{\beta^{-1}\zeta^{-1}}  \WW^2_T \ll \frac{1}{2} \int_0^T \left( \U(t,\uu{X}^1_t) + \U(t,\uu{X}^2_t) \right) \, dt,
\end{equation}
 \end{itemize}
to obtain the following equations:
\begin{equation}\label{eq:X_lag}
 \left\lbrace
 \begin{aligned}
 &\dps{d \XX_t = \nabla \U(t, \uu{R}_t) \XX_t \, dt  - 2\zeta^{-1}
 \FF(\XX_t) \, dt +  \sqrt{4 \beta^{-1}\zeta^{-1}} d \WW_t},\\
 &\dps{d \uu{R}_t =  \U(t,\uu{R}_t) \, dt}.
 \end{aligned}
 \right.
\end{equation}
The Eulerian version of the Lagrangian formulation~\eqref{eq:X_lag} is
\begin{equation}\label{eq:X}
 \begin{aligned}
 d \XX_t(\uu{x}) +&  \U(t,\uu{x}). \nabla \XX_t(\uu{x})\, dt \\
&=  \nabla \U(t,\uu{x}) \XX_t(\uu{x}) \, dt  - 2\zeta^{-1}
 \FF(\XX_t(\uu{x})) \, dt +  \sqrt{4 \beta^{-1}\zeta^{-1}} d \WW_t.
\end{aligned}
\end{equation}
In~\eqref{eq:X_lag}, the process $\XX_t$ is (implicitly) indexed by $\uu{R}_0$ which is the initial position of the dumbbell in the solvent. In~\eqref{eq:X}, $\XX_t(\uu{x})$ depends on the position at time $t$ of the dumbbell in the solvent. Notice that the random variable $\XX_t(\uu{x})$ is a function of the time $t$, the position $\uu{x}$, and the probability variable $\omega$.

\subsubsection{Discussion of the modelling}

A few remarks are in order concerning the modelling procedure described above. Some of these remarks are not specific to the modelling of polymeric fluids, but concern more generally molecular dynamics.

First, concerning the coarse-graining procedure in the solvent at rest ($\U=0$), we have derived the effective potential so that the canonical measure $\exp(-\beta \Pi(\xx)) d \xx$ is consistent with the original full canonical measure $\mu(d\qq)$, at the thermodynamic equilibrium (in the sense of~\eqref{eq:FE_def_}). But then, we have used this potential to derive {\em a dynamics} on the end-to-end vector:
$$d\XX_t = - 2 \zeta^{-1} \nabla \Pi (\XX_t) \, dt + \sqrt{4 \beta^{-1} \zeta^{-1}} \, d\WW_t.$$
It is easy to check that this dynamics indeed leaves invariant the canonical measure $\exp(-\beta \Pi(\xx)) d \xx$ (it is in some sense consistent in terms of the thermodynamic equilibrium) but a natural question is then the relation between the dynamics $(\xi(\qq_t))_{t \ge 0}$ and $(\XX_t)_{t \ge 0}$. Are the time-correlation functions correctly approximated using the coarse-grained dynamics $\XX_t$~? Are the transition times between metastable states correctly captured~? This is the subject of current researches~\cite{legoll-lelievre-10}.

Second, the coarse-graining procedure is even more questionable for a solvent in motion. A first question is: what are the correct thermalizing terms in such a situation~? In other words, what are the Langevin equations for a particle within a moving solvent ? The above approach considers a friction force involving the relative velocities $-\zeta ( M^{-1} \PP_t - u(t,\QQ_t) )$. Can this be justified in a rigorous way ? A second question is the relevance of the effective potential $\Pi$ originally derived in the solvent at rest for a setting where the solvent moves. For example, the image by $\xi$ of the stationary state for the original Langevin dynamics on the finest model in a velocity field $\U \neq 0$ is certainly not the stationary state obtained for the coarse-grained model~\eqref{eq:X}. In other words, the coarse-graining procedure (using the free energy) is sensitive to velocity of the solvent.

Third, the expansion on the velocity field~\eqref{eq:exp_u} introduces the gradient of the velocity field in the model while it was originally absent. This leads to some mathematical difficulties, since it requires some smoothness on the velocity field.

Fourth, neglecting the noise on $\uu{R}_t$ (see~\eqref{eq:small_noise}) also has some consequences on the mathematical analysis. Keeping the noise term would amount to adding some diffusion in space on the dumbbell position, which would dramatically simplify the mathematical analysis of the system.

Finally, we would like to mention that we have described above a suitable model for dilute solution of polymers (in particular direct interaction of polymer chains are neglected). Similar descriptions have been used to model:
\begin{itemize}
\item rod-like polymers and liquid crystals~\cite{doi-edwards-88,oettinger-95},
\item polymer melts~\cite{doi-edwards-88,oettinger-95},
\item concentrated suspensions~\cite{hebraud-lequeux-98},
\item blood~\cite{owens-06}.
\end{itemize}
Such kinetic models are thus useful in many contexts.

%% 1D %%

\subsection{The micro-macro model for polymeric fluids}

\subsubsection{Kramers expression for the stress tensor}

In this section, we would like to couple equations~\eqref{eq:momentum} at the macroscopic level to equations~\eqref{eq:X} at the microscopic level. This requires an expression for the extra-stress tensor $\ttau$ as a function of the polymer chains configurations.

Such an expression is standard in the context of molecular dynamics. It carries several names: Irving-Kirkwood formula, virial stress or Kramers expression. It writes, in our context:
\begin{equation}\label{eq:tau}
 \ttau(t,\uu{x})=n_p \Big( - k T \I + \E \left( \XX_t(\uu{x}) \otimes \FF(\XX_t(\uu{x})) \right) \Big),
\end{equation}
where $n_p$ denotes the polymer concentration (which is supposed to be fixed and constant in this model), $\E$ is the expectation (integral with respect to the probability variable $\omega$), and $\otimes$ denotes a tensor product (for two vectors $\uu{u}$ and $\uu{v}$, $\uu{u} \otimes \uu{v}$ is a matrix with elements $(u_i v_j)$). This formula can for example be derived starting from the definition of the stress in terms of the force exerted on a plane within the fluid, see~\cite{oettinger-95}.

\subsubsection{The coupled system}

Collecting equations~\eqref{eq:momentum}, \eqref{eq:X} and~\eqref{eq:tau},  the complete coupled system 
$$
\left\lbrace
\begin{aligned}
&\rho \left( \frac{\partial\U }{\partial t}  + \U.\nabla \U \right) = - \nabla p + \eta
\Delta \U + { \div ( \ttau)},\\
&\div(\U)=0, \\
&\ttau=n_p \Big( - k T \I + \E \left( \XX_t \otimes \FF(\XX_t) \right)\Big),\\
&d\XX_t + { \U .\nabla_{\uu{x}} \XX_t} \, dt =\left( { \nabla  \U \XX_t} - \frac{2}{\zeta} \FF(\XX_t)
\right) \,dt + \sqrt{\frac{4 k T} {\zeta}} d \WW_t,
\end{aligned}
\right.
$$
is obtained. Once non-dimensionalized, it writes:
\begin{equation}\label{eq:micmac}
\left\lbrace
\begin{aligned}
&\Rey \left(\frac{\partial \U}{\partial t} + \U \cdot \nabla \U \right)   =
- \nabla p + (1-\epsilon) \Delta \U  + \div(\ttau), \\
&\div(\U) = 0,\\
&\displaystyle\ttau = \frac{\epsilon}{\We}( \E(\XX_t \otimes \FF(\XX_t)) - \I),\\
&d\XX_t + \U \cdot \nabla_{\uu{x}} \XX_t \, dt = \left( \nabla \U \cdot \XX_t -
  \frac{1}{2 { \We}} \FF(\XX_t) \right) dt +\frac{1}{\sqrt{{ \We}}} d\WW_t,
\end{aligned}
\right.
\end{equation}
with the following non-dimensional numbers:
$$\Rey = \frac{\rho U L}{\eta} \textrm{, }
\We = \frac{\lambda U}{L} \textrm{, }
\epsilon =  \frac{\eta_p}{\eta}$$
where $L=\sqrt{\frac{k T}{H}}$ is chosen as the characteristic length scale,
 $\lambda=\frac{\zeta}{4H}$ is the relaxation time of the polymer chains,
$\eta_p=n_p k T \lambda$ is the viscosity associated to the polymers, and
$U$ is a characteristic velocity. For the Hookean model,
\begin{equation}\label{eq:hook}
\FF(\XX)=\XX \text{ and } \Pi(\XX)=\frac{|\XX|^2}{2},
\end{equation}
and for the FENE model,
\begin{equation}\label{eq:FENE}
\FF(\XX)=\frac{\XX}{1 - |\XX|^2 / b} \text{ and } \Pi(\XX)= - \frac{b}{2} \ln \left(1 - |\XX|^2 / b\right).
\end{equation}
Of course, the system of equations~\eqref{eq:micmac} needs to be supplied with initial conditions and boundary conditions. In the following, we denote ${\mathcal D}$ the physical domain occupied by the fluid.

The micro-macro model~\eqref{eq:micmac} can be rewritten equivalently as:
\begin{equation}\label{eq:micmac-FP}
\left\{
\begin{aligned}
&\Rey \left(\frac{\partial \U}{\partial t} + \U \cdot \nabla \U \right) = -\nabla p + (1- \epsilon) \Delta \U + \div (\ttau), \\
&\div (\U) =0,\\
&\ttau = \frac{\epsilon}{\We} \left( - \I + \int (\XX \otimes \FF(\XX)) \psi \, d \XX \right), \\
& \frac{\partial \psi}{\partial t}  + \U \cdot \nabla_{\xx} \psi = - \div_{\XX} \left( \left(\nabla \U \XX - \frac{1}{2 \We} \FF (\XX) \right) \psi \right) + \frac{1}{2 \We} \Delta_\XX \psi,
\end{aligned}
\right.
\end{equation}
where $\psi(t,\xx,\XX) \, d\XX$ is the law of the random variable $\XX_t(x)$. The partial differential equation on $\psi$ is called the {\em Fokker-Planck equation}. It is derived from the stochastic differential equation on $\XX_t$ in~\eqref{eq:micmac}.

\subsubsection{Relation to macroscopic models}

It is interesting to notice at this stage that the Hookean dumbbell model is equivalent to the Oldroyd-B model~\eqref{eq:OB}. Indeed, if $\FF(\XX)=\XX$, it is easy to check that $\ttau$ satisfies:
\begin{equation}\label{eq:OB-nd}
\displaystyle{\frac{\partial \ttau}{\partial t} + \U.\nabla
    \ttau =  \nabla \U \ttau + \ttau (\nabla \U)^T +
  \frac{\epsilon}{\We} (\nabla \U + (\nabla \U)^T) - \frac{1}{\We} \ttau.}
\end{equation}
\begin{rem}\label{rem:UCM}
The derivative  $\frac{\partial \uu{\tau}}{\partial t} + \U\cdot \nabla \uu{\tau} - \nabla    \U  \uu{\tau} - \uu{\tau}(\nabla   \U)^T$ is called the {\em upper convected derivative}. It is known to be a frame invariant derivative. Other frame invariant derivatives exist~\cite{BCAH-87-1,BCAH-87-2}, like for example the corotational derivative
\begin{equation}\label{eq:corot_der}
\frac{\partial \ttau}{\partial t} + \U.\nabla
    \ttau - { W(\U)} \ttau - \ttau { W(\U)}^T,
\end{equation} where $W(\U)=\frac{\nabla \U - \nabla \U^T}{2}$. The upper convected derivative is the frame invariant derivative which naturally appears in macroscopic models derived from microscopic models.
\end{rem}

There is no known macroscopic equivalent to the FENE model. However, the {\em closure approximation}~\cite{peterlin-66,bird-dotson-johnson-80}
$\E\left( \frac{\XX_t}{ 1 - |\XX_t|^2 / b} \right) \simeq \frac{\E(\XX_t)}{ 1 - \E(|\XX_t|^2) / b}$ 
leads to the FENE-P model, which is another classical macroscopic model (where $\uu{A}=\E(X_t \otimes X_t)$):
\begin{equation}\label{eq:FENE-P}
\begin{aligned}
\ttau&=\frac{\epsilon}{\We} \left( \frac{ \uu{A} }{1 -
    \tr(\uu{A})/b} - \I \right),\\
\frac{\partial \uu{A}}{\partial t} + \U \cdot\nabla \uu{A} &= \nabla \U  \uu{A}
+\uu{A}  \nabla \U^T - \frac{1}{\We} \frac{ \uu{A} }{1 -
    \tr(\uu{A})/b} + \frac{1}{\We} \I.
\end{aligned}
\end{equation}

\subsection{Micro-macro models {\em versus} macroscopic models}\label{sec:mic_vs_mac}

The superiority of the micro-macro approach (system~\eqref{eq:micmac}) compared to a classical macroscopic approach relies in the quality of the modelling. The approach enables numerical exploration of the link between microscopic properties (the polymer chains) and macroscopic properties (the rheology of the fluid). This is crucial for deriving more predictive models in order to test the behaviour of a known fluid in new regimes, or to explore the properties of a fluid that has not been synthetized yet. The microscopic features of a polymer chain (and more generally of the microstructures within a fluid) are indeed typically well known, but the way these microstructures affect the rheology of the fluid is very difficult to predict. Micro-macro modelling is a way to answer this question using numerical experiments.

However, micro-macro approaches remain so far academic tools because of their very high computational cost due to the necessary introduction of additional microscopic variables. Let us also mention that, in our context of polymeric fluids, these models share the numerical difficulties of the classical macroscopic approaches. In particular, instabilities occur in some geometries when the Weissenberg number becomes large (this is the so-called {\em High Weissenberg Number Problem}, see Section~\ref{sec:num_gen}).

These considerations are summarized in Table~\ref{tab:micmac_vs_macmac}.

Because of the computational costs mentioned above, it seems natural to try and design some
numerical methods that combine the macroscopic and the micro-macro
approaches. The  macro-macro model is used where the flow is
 simple, and the detailed micro-macro model is used elsewhere. The idea  of adaptive modelling based on {\it a posteriori}
modelling error  analysis (see J.T.~Oden and K.S.~Vemaganti~\cite{oden-vemaganti-00},
 J.T.~Oden and S.~Prudhomme~\cite{oden-prudhomme-02} or M.~Braack and
 A.~Ern~\cite{braack-ern-03}) has
 been adapted in this context in a preliminary work by A.~Ern
 and T.~Lelièvre~\cite{ern-lelievre-07}.

\begin{table}
\begin{tabular}{c|c|c|c}
\cline{2-4}&MACRO& \multicolumn{2}{c}{MICRO-MACRO}\\[4pt]
\hline modelling capabilities & low & \multicolumn{2}{c}{high} \\[8pt]
\hline
current utilization & industry & \multicolumn{2}{c}{laboratories} \\[8pt]
\hline&   & discretization  & discretization \\
&   & by Monte Carlo  & of Fokker-Planck \\[4pt]
\hline computational cost & low & high & moderate \\[8pt]
\hline computational bottleneck & HWNP  &  variance, HWNP & dimension, HWNP \\[8pt]
\hline
\end{tabular}
\caption{Pros and cons for the macro-macro and micro-macro
  approaches.}\label{tab:micmac_vs_macmac}
\end{table}

\section{Mathematical analysis}\label{sec:math}

\subsection{Generalities}\label{sec:math_gen}

The main difficulties for the mathematical analysis of the micro-macro systems~\eqref{eq:micmac} (or~\eqref{eq:micmac-FP}) come (as always) from the nonlinear terms in the equations namely the transport terms $\U \cdot \nabla \U$, $\U \cdot \nabla_x \XX_t \, dt$ or $\U \cdot \nabla_x \psi$, and the coupling terms $\nabla \U \XX_t \, dt$ or $\div_{\XX}( \nabla \U \XX \psi )$. There exist similar difficulties for the macroscopic models, like the Oldroyd-B model. Until recently, only local-in-time existence and uniqueness results were available for such models. Recent works~\cite{masmoudi-10-a,masmoudi-10-b} by N. Masmoudi (following the earlier work~\cite{lions-masmoudi-00} by P.L. Lions and N. Masmoudi) recently changed the landscape.

The most recent results are based on {\em a priori} entropy estimates, and fine results on the propagation of compactness in the problem.

\begin{rem}
The separation between the coupling terms and the transport terms mentioned above is somehow misleading. Actually, all these terms can be seen as transport terms. Let us for example discuss the upper convected derivative which appears in the Oldroyd-B model~\eqref{eq:OB-nd}. One can check that if $\uu{y}(t,\uu{y}_0)$ satisfies:
$$\left\{
\begin{aligned}
\frac{d}{dt} \uu{y}(t,\uu{y}_0) &= \U(t,\uu{y}(t,\uu{y}_0)),\\
\uu{y}(0,\uu{y}_0)&=\uu{y}_0,
\end{aligned}
\right.
$$
then $\uu{G}(t,y(t,\uu{y}_0))=\frac{\partial}{\partial \uu{y_0}} y(t,\uu{y}_0)$ satisfies
$$\frac{\partial \uu{G}}{\partial t} + \U \cdot \nabla \uu{G} = \nabla \U \uu{G}$$ and
$\uu{\sigma}(t,y)=\uu{G}(t,y) \uu{\sigma}_0 \uu{G}^T(t,y)$ satisfies
 $$\frac{\partial \uu{\sigma}}{\partial t} + {\U.\nabla
    \uu{\sigma}} - {\nabla \U \uu{\sigma} - \uu{\sigma} (\nabla \U)^T} = 0.$$
 Likewise, for the Fokker-Planck equation in~\eqref{eq:micmac-FP}, one can check that:
\begin{align*}
&\frac{d}{dt} \Big(\psi(t,\uu{y}(t,\uu{y}_0),\uu{G}(t,\uu{y}_0) \XX) \Big)\\
&= \Big(\frac{\partial \psi}{\partial t} + \U\cdot\nabla_{\uu{x}}\psi +
     \div_\XX\,\left(\nabla_{\uu{x}}\U\,\XX \psi \right) \Big)(t,\uu{y}(t,\uu{y}_0),\uu{G}(t,\uu{y}_0) \XX).
\end{align*}
This rewriting of these terms along characteristics is well-known in the literature, but it does not seem to be useful as such for theoretical nor numerical purposes (see however~\cite{lee-xu-06}).
\end{rem}

\subsection{{\em A priori} estimates, entropy and longtime behaviour}

A fundamental tool to analyze a model is a so-called {\em a priori} estimate which gives some bound on the physical quantities. These estimates are useful to prove existence results and to analyze the longtime behaviour of the system. In the following, unless explicitly stated, we assume homogeneous boundary conditions: $\U=0 \text{ on } \mathcal D$. Notice that the {\em a priori} estimates we derive and the results on the longtime behaviour we obtain below assume the existence of a smooth solution.

\subsubsection{A bad energy estimate}

A first natural energy estimate is obtained by the following procedure. First, multiplying the momentum equation by $\U$ and integrating, one obtains:
\begin{equation}\label{eq:Ec}
\frac{\Rey}{2} \frac{d}{dt} \int_{\mathcal D} |\U|^2 + (1 - \epsilon) \int_{\mathcal D} |\nabla \U|^2 = - \int_{\mathcal D} \ttau : \nabla \U.
\end{equation}
Using It\^o calculus, the time derivative of the potential energy stored in the springs is:
\begin{equation}\label{eq:Ep}
\frac{d}{dt} \int_{\cal D} \E(\Pi(\XX_t)) + \frac{1}{2 \We} \int_{\cal D} \E(|\FF(\XX_t)|^2) =  \int_{\cal D} \E ( \FF(\XX_t) \cdot \nabla \U \XX_t ) + \frac{1}{2 \We}  \int_{\cal D}
\Delta \Pi(\XX_t).
\end{equation}
We recall that $\Pi$ is the potential associated to the entropic force $F$ ($F=\nabla \Pi$). Using the fact that $\ttau=\frac{\epsilon}{\We}\left(- \I + \E(\XX_t \otimes F(\XX_t))\right)$ and the the potential $\Pi$ is radial (which implies that $\ttau$ is a symmetric tensor), one can check that $\ttau : \nabla \U = \E ( \FF(\XX_t) \cdot \nabla \U \XX_t )$, and thus one obtains from~\eqref{eq:Ec} and~\eqref{eq:Ep}:
\begin{align}
\frac{d}{dt} \left( \frac{\Rey}{2} \int_{\cal D} |\U|^2 \right. & \left. + \frac{\epsilon}{\We}  \int_{\cal D}
\E(\Pi(\XX_t)) \right) + (1-\epsilon)  \int_{\cal D} |\nabla \U|^2  + \frac{\epsilon}{2 \We^2}  \int_{\cal D} \E(|\FF(\XX_t)|^2) \nonumber \\
 &= \frac{\epsilon}{2 \We^2} \int_{\cal D}
\E(\Delta \Pi(\XX_t)).\label{eq:estim_NRJ}
\end{align}
Note that~\eqref{eq:estim_NRJ} also writes in terms of the density probability $\psi$ (this estimate can actually be obtained directly from~\eqref{eq:micmac-FP}):
\begin{align}
\frac{d}{dt} \left( \frac{\Rey}{2} \int_{\cal D} |\U|^2 \right. & \left. + \frac{\epsilon}{\We}  \int_{\cal D} \int_{\R^d} \Pi \, \psi \right) + (1-\epsilon)  \int_{\cal D} |\nabla
\U|^2  + \frac{\epsilon}{2 \We^2}
  \int_{\cal D} \int_{\R^d} |\FF|^2 \, \psi \nonumber \\
 &= \frac{\epsilon}{2 \We^2} \int_{\cal D} \int_{\R^d}
\Delta \Pi \, \psi,\label{eq:estim_NRJ-FP}
\end{align}
where $\int_{\R^d}$ is an integral over the microscopic variable $\XX$ ($d$ is the dimension of the ambient space: ${\mathcal D} \subset \R^d$).

The potential $\Pi$ is strongly convex (see~\eqref{eq:hook} and~\eqref{eq:FENE}) so that the right-hand sides of~\eqref{eq:estim_NRJ} and~\eqref{eq:estim_NRJ-FP} are positive. This shows that these estimates, although informative on a finite time interval, will not be useful to study the longtime behaviour. Since no external forces apply on the system, a return to equilibrium is expected: in the longtime limit, the velocity should converge to $\U=0$ and the law of $\XX_t$ should converge to $Z^{-1} \exp (- \Pi)$. One expects the system to be dissipative and this does not seem to translate on~\eqref{eq:estim_NRJ}. The question is then how to eliminate the term in the right-hand side. For this purpose, we will need to introduce an entropy.

\subsubsection{Entropy estimates: basic facts}

To obtain an energy estimate useful to study the longtime behaviour, we need to consider the micro-macro model~\eqref{eq:micmac-FP} written in terms of the density $\psi$ rather than its stochastic counterpart~\eqref{eq:micmac}.

We first recall the principle of {\em entropy methods}~\cite{arnold-markowich-toscani-unterreiter-01,ABC-00} on the Fokker-Planck equation itself, the velocity field being given. Consider the linear problem:
\begin{equation}\label{eq:FP}
\frac{\partial \psi}{\partial t} =
     \div_\XX\left(\left(- \uu{\kappa} \XX + \frac{1}{2 \We} \nabla \Pi(\XX) \right)\psi
     \right)
     +\frac{1}{2 \We }\,\Delta_\XX \psi,
\end{equation}
where $\psi$ is only a function of the time variable $t$ and of the microscopic variable $\XX$, and $\uu{\kappa}$ is a given tensor, which is the gradient of the velocity field. A now classical approach to study the longtime behaviour of such an equation consists in introducing the {\em relative entropy}
\begin{equation}\label{eq:rel_ent}
H(\psi | \psi_\infty) = \int_{\R^d}   \ln \left(\frac{\psi}{\psi_\infty}\right) \psi
\end{equation}
 of $\psi$ with respect to a stationary state $\psi_\infty$  of~\eqref{eq:FP}. Notice that  $H \ge 0$ and that $H=0$ if and only if $\psi = \psi_\infty$. Let us recall the {\em Csiszar-Kullback inequality}
\begin{equation}\label{eq:CK}
\int_{\R^d}|\psi - \psi_\infty| \le \sqrt{ 2 H(\psi | \psi_\infty)},
\end{equation}
which shows that the entropy~\eqref{eq:rel_ent} provides a control on the $L^1$-norm of $\psi-\psi_\infty$.
Using the fact that $\psi_\infty$ is a stationary solution to~\eqref{eq:FP}, one can check that
\begin{equation}\label{eq:entropie}
\frac{d H(\psi | \psi_\infty)} {dt} = -\frac{1}{2 \We} \int
 \left|\nabla
 \ln \left(\frac{\psi}{\psi_\infty}\right) \right|^2 \psi.
\end{equation}
The right-hand side is non positive, and the entropy $H$ thus decreases in time.

In order to obtain the convergence of $H$ to $0$ with a specific rate of convergence (and thus the convergence of $\psi$ to $\psi_\infty$ using~\eqref{eq:CK}), one then needs a {\em functional inequality} on the stationary measure $\psi_\infty$, called a {\em logarithmic Sobolev inequality}. It is said that $\psi_\infty$ satisfies a logarithmic Sobolev inequality if there exists a constant $\rho>0$ such that
\begin{equation}\label{eq:ineg_func}
\int \ln \left(\frac{\phi}{\psi_\infty}\right) \phi \leq \frac{1}{2 \rho} \int
 \left|\nabla
 \ln \left(\frac{\phi}{\psi_\infty}\right) \right|^2 \phi,
\end{equation}
for all non-negative function $\phi$ with integral $1$.
Inserting~\eqref{eq:ineg_func} in~\eqref{eq:entropie} and using the Gronwall Lemma, one obtains the exponential convergence: $\forall t \ge 0$,
$$H(\psi(t,\cdot) | \psi_\infty) \le H(\psi(0,\cdot) | \psi_\infty) \exp \left(- \frac{\rho}{ \We} t \right).$$
We are left with the question of proving a logarithmic Sobolev inequality~\eqref{eq:ineg_func} for $\psi_\infty$. This is a delicate question (especially when no analytical expression for $\psi_\infty$ is available) and a very active subject of research. A related issue is the identification of the spectrum of a Fokker-Planck operator with a drift term that is not in gradient form.

% We start with a formal definition.
% \begin{defi}\label{def:LSI}
% A probability measure  $\nu$ satisfies a logarithmic Sobolev inequality with % constant $\rho>0$ (in short LSI($\rho$)) if and only if, for all probability % measure $\mu$ such that $\mu$ is absolutely continuous with respect to $\nu$ % (denoted $\mu \ll \nu$ in the following)
% $$H(\mu | \nu) \le \frac{1}{2 \rho} I( \mu | \nu)$$
% where $H(\mu | \nu) = \dps{\int \ln \left( \frac{d \mu}{ d \nu} \right) d\mu}$ % is the relative entropy of the measure $\mu$ with respect to the measure $\nu$ % and $I(\mu | \nu) = \dps{\int \left| \nabla \ln \left( \frac{d \mu}{ d \nu} % \right)\right|^2 d\mu}$ is the Fisher information of the measure $\mu$ with % respect to the measure $\nu$.
% \end{defi}
% With a slight abuse of notation, we will denote $H(\psi| \phi)$ the relative % entropy of a probability measure $\psi(\XX) \, d\XX$ with respect to a % probability measure $\phi(\XX) \, d\XX$.

Let us recall some standard criteria to obtain a logarithmic Sobolev inequality. There currently exist three basic tools to prove that a probability measure satisfies a logarithmic Sobolev inequality. The first one is the {\em Bakry-Emery criterion}:
\begin{prop}\label{prop:BE}
Let $\Pi$ be an $\alpha$-convex function, that is a function such that for all vector $\XX$, $\XX^T \nabla^2 \Pi \XX \ge \alpha |\XX|^2$ (where $\nabla^2 \Pi$ is the Hessian of $\Pi$). Then, the density probability $\psi_\infty \varpropto \exp(-\Pi)$ satisfies a logarithmic Sobolev inequality with constant $\rho \ge \alpha$.
\end{prop}
The second one is a perturbative result due to Holley and Stroock:
\begin{prop}\label{prop:HS}
Let $\Pi$ be a function such that the probability density $\psi_\infty \varpropto \exp(-\Pi)$ satisfies a logarithmic Sobolev inequality with constant $\rho$. Let $\tilde\Pi$ be a bounded function, and let us introduce the probability density $\tilde \psi_\infty \varpropto \exp(-\Pi + \tilde{\Pi})$. Then $\tilde \psi_\infty$ also satisfies a logarithmic Sobolev inequality with constant $\tilde\rho \ge \rho \exp(- \osc (\tilde{\Pi}))$ where $\osc (\tilde{\Pi})=\sup \tilde{\Pi} - \inf \tilde{\Pi}$.
\end{prop}
The third tool will be less useful in our specific context but is crucial in other fields. It relates to tensorization of probability measures. If $(\nu_i)_{1 \le i \le I}$ are probability measures that satisfy logarithmic Sobolev inequalities respectively with constant $\rho_i$, then $\nu_1 \otimes \ldots \otimes \nu_I$ also satisfies a logarithmic Sobolev inequality, with constant $\min(\rho_1, \ldots, \rho_I)$. This result for a collection of $I$ independent random variables admits generalizations when some correlations are introduced, see~\cite{otto-reznikoff-07,grunewald-otto-villani-westdickenberg-09,lelievre-09}.

\medskip
Going back to the Fokker-Planck equation~\eqref{eq:FP}, one can check~\cite{jourdain-le-bris-lelievre-otto-06} that:
\begin{itemize}
\item If $\uu{\kappa}$ is zero or skew-symmetric, $\psi$ converges exponentially fast to $\psi_\infty \varpropto \exp(-\Pi)$ (since $\Pi$ is $\alpha$-convex).
\item If $\uu{\kappa}$ is symmetric, an analytical expression for the stationary state is again available: $\psi_\infty \varpropto \exp(-\Pi + \We X^T \uu{\kappa} X)$, and thus one obtains exponential convergence for the FENE model for any values of the Weissenberg number $\We$ (using the fact that in the FENE model, $\XX$ leaves in a bounded domain), and for the Hookean model under the assumption that the eigenvalues of $\uu{\kappa}$ are strictly smaller than $\frac{1}{2\We}$.
\item For a general tensor $\uu{\kappa}$ and using Holley-Stroock criterion, exponential decay to equilibrium is obtained if there exists a stationary state $\psi_\infty$ that satisfies $\|\ln(\psi_\infty \exp(\Pi))\|_{L^\infty} < \infty$. This can be proved for the FENE model under the assumption that the symmetric part of $\uu{\kappa}$ is sufficiently small (see~\eqref{eq:estim_psi} below).
\end{itemize}

\subsubsection{Entropy estimates for the coupled micro-macro system: $\U=0$ on $\partial {\mathcal D}$}\label{sec:estim_ent}

We now return to the coupled problem~\eqref{eq:micmac-FP}, and first assume for simplicity that $\U=0$ on the boundary $\partial {\mathcal D}$ (no external forcing). In this case, a natural and expected stationary state is $\U_\infty=0$ and $\psi_\infty \varpropto \exp(-\Pi)$. Let us reconsider the  energy estimate~\eqref{eq:estim_NRJ-FP}, and introduce the entropy instead of the energy stored in the springs. This consists in replacing in~\eqref{eq:estim_NRJ-FP} the term $\int_{\mathcal D} \int_{\R^d} \Pi \, \psi$ by $\int_{\mathcal D} \int_{\R^d} \ln \left( \frac{\psi}{\psi_\infty} \right) \psi$ (which is, up to an additive constant, $\int_{\mathcal D} \int_{\R^d} \Pi \, \psi + \int_{\mathcal D} \int_{\R^d}  \psi \ln \psi$). Then, a straightforward calculation yields:
\begin{equation}\label{eq:Ep2}
\frac{d}{dt} \int_{\mathcal D} \int_{\R^d} \ln \left( \frac{\psi}{\psi_\infty} \right) \psi + \frac{1}{2 \We} \int_{\mathcal D} \int_{\R^d} \left| \nabla_\XX \ln \left( \frac{\psi}{\psi_\infty} \right) \right|^2 \psi=  \int_{\mathcal D} \int_{\R^d} \FF(\XX) \cdot \nabla \U \XX \, \psi,
\end{equation}
instead of~\eqref{eq:Ep}.
By linear combination of~\eqref{eq:Ec} and~\eqref{eq:Ep2}, one obtains (compare with~\eqref{eq:estim_NRJ-FP}):
\begin{equation}\label{eq:estim_ent}
\begin{aligned}
\frac{d}{dt}& \left( \frac{\Rey}{2} \int_{\cal D} |\U|^2  + \frac{\epsilon}{\We}  \int_{\cal D} \int_{\R^d} \ln \left( \frac{\psi}{\psi_\infty} \right)  \psi \right) \\
&+ (1-\epsilon)  \int_{\cal D} |\nabla \U|^2  + \frac{\epsilon}{2 \We^2}
  \int_{\cal D} \int_{\R^d}  \left| \nabla_\XX \ln \left( \frac{\psi}{\psi_\infty} \right) \right|^2  \psi  =0.
\end{aligned}
\end{equation}
The energy $\frac{\Rey}{2} \int_{\cal D} |\U|^2  + \frac{\epsilon}{\We}  \int_{\cal D} \int_{\R^d} \Pi \,  \psi$ is not a decreasing function  (see~\eqref{eq:estim_NRJ-FP}), but the free energy $\frac{\Rey}{2} \int_{\cal D} |\U|^2  + \frac{\epsilon}{\We}  \int_{\cal D} \int_{\R^d} \ln \left( \frac{\psi}{\psi_\infty} \right)  \psi$ does indeed decrease in time (see~\eqref{eq:estim_ent}). The dissipative nature of the system is obvious on~\eqref{eq:estim_ent}.

Using the logarithmic Sobolev inequality satisfied by $\psi_\infty$ (which holds since $\Pi$ is $\alpha$-convex), and the Poincaré inequality
\begin{equation}\label{eq:PI}
\int_{\mathcal D} |\U|^2 \le C_{\rm PI}({\mathcal D}) \int_{\mathcal D} |\nabla \U|^2,
\end{equation}
one obtains the convergence of $(\U,\psi)$ to the equilibrium $(\U_\infty,\psi_\infty) = (0, Z^{-1} \exp(-\Pi))$ at exponential rate.

Two remarks are in order. First, in order to write the free energy of the system, we have considered the Fokker-Planck version~\eqref{eq:micmac-FP} of the micro-macro system and not the stochastic formulation~\eqref{eq:micmac}. It is unclear how to obtain a similar estimate only in terms of the random variable $\XX_t(\xx)$ (which would be useful for analyzing the stability of numerical schemes based on the stochastic formulation~\eqref{eq:micmac}, for example). Second, on the linear problem~\eqref{eq:FP}, it is actually possible to analyze the longtime behaviour using many entropy functionals of the form $$H(\psi | \psi_\infty) = \int_{\R^d}   h \left(\frac{\psi}{\psi_\infty}\right) \psi_\infty,$$
where $h$ is a regular scalar convex function such that $h(1)=h'(1)=0$ see~\cite{arnold-markowich-toscani-unterreiter-01}. Classical examples include $h(x)=x \ln (x) - x + 1$ (classical "physical" entropy, related to logarithmic Sobolev inequalities) and $h(x)=(x-1)^2$ ($\chi^2$-distance, related to Poincaré inequalities). On the nonlinear problem~\eqref{eq:micmac-FP}, $h(x)=x \ln (x) - x + 1$ which yields the entropy~\eqref{eq:rel_ent} introduced above is seemingly the only choice that leads to the dissipative structure~\eqref{eq:estim_ent}. This is a generic observation: for many nonlinear Fokker-Planck equations (in the sense of MacKean), the correct entropy seems to be the "physical" relative entropy~\eqref{eq:rel_ent}. This may be related to the fact that this entropy naturally passes to the limit when considering discretizations using interacting particle systems (extensivity property).

\subsubsection{Entropy estimates for the coupled micro-macro system: $\U \neq 0$ on $\partial {\mathcal D}$}

A natural question is to generalize the result of the previous section to a system with non-zero forcing. This falls into the general question of convergence to {\em non-equilibrium steady states}. The question is known to be much more difficult than the study of equilibrium steady states (namely steady-stated without external forcing, at the thermodynamic equilibrium).

In our context, we consider a forced system through the boundary:  $\U \neq 0$ on $\partial {\mathcal D}$. In such a situation, a stationary state  $(\U_\infty,\psi_\infty)$ is not the equilibrium state $(0, Z^{-1} \exp(-\Pi))$, but is simply defined as a stationary state to~\eqref{eq:micmac-FP} satisfying the boundary conditions. For simplicity and without loss of generality, let us assume in this section that: $\Rey=1/2$, $\We=1$ and $\epsilon=1/2$.

Mimicking the reasoning of the previous section we obtain, instead of~\eqref{eq:estim_ent},
\begin{align}
 & \frac{d}{dt}\left( \frac{1}{2} \int_{{\cal D}} |\uu{\ou}|^2 +
    \int_{{\cal D}}\int_{ \R^d}  \ln\left(
      \frac{\psi}{\psi_\infty}\right) \psi \right) + \int_{{\cal D}} |\nabla \uu{\ou}|^2  +\frac{1}{2} \int_{{\cal D}} \int_{\R^d}  \left| \nabla_\XX
 \ln \left( \frac{\psi}{\psi_\infty} \right) \right|^2 \psi  \nonumber \\
&= - \int_{{\cal D}} \uu{\ou} \cdot \nabla \U_\infty \uu{\ou}  - \int_{{\cal D}} \int_{\R^d} \uu{\ou}\cdot\nabla_{\xx} (\ln\psi_\infty)
\opsi \nonumber \\
&\quad - \int_{{\cal D}}\int_{\R^d} \left ( \nabla_{\XX}
(\ln \psi_\infty) +  \nabla \Pi (\XX) \right) \cdot \nabla \uu{\ou} \XX  \,
\opsi , \label{eq:estim_ent_noneq}
\end{align}
where $\uu{\ou}(t,\xx) =\U(t,\xx) - \U_\infty(\xx)$ and $\opsi(t,\xx,\XX)=\psi(t,\xx,\XX)-\psi_\infty(\xx,\XX)$. Compared to the situation $\U=0$ on $\partial \Omega$, there are two new difficulties: (i) new terms in the right-hand side of~\eqref{eq:estim_ent_noneq} appear and  have to be controlled; (ii) one needs to prove that $\psi_\infty$ satisfies a logarithmic Sobolev inequality.

For the FENE model, it is actually possible to obtain exponential decay to equilibrium. A typical situation where a proof can be performed is the following (we refer to~\cite{jourdain-le-bris-lelievre-otto-06} for more details). We assume that the boundary conditions are compatible with a homogeneous stationary state: $\U_\infty(\xx) = \uu{\kappa} \xx$ (where $\uu{\kappa}$ is a trace free matrix). Then, $\psi_\infty$ is assumed not to depend on the space variable $\xx$. Thus, the second term in the right-hand side of~\eqref{eq:estim_ent_noneq} is zero. The first term can be controlled by the left-hand side if $\uu{\kappa}$ is sufficiently small, using a Poincaré inequality on $\uu{\ou}$. To have an estimate of the last term, as well as to prove a logarithmic Sobolev inequality for $\psi_\infty$ (using Proposition~\ref{prop:HS}), one needs to control $\|\nabla_\XX \ln( \psi_\infty \exp(\Pi) )\|_{L^\infty}$. This can be achieved under an additional assumption on~$\uu{\kappa}$. More precisely, one can prove that if
$$|\uu{\kappa}^s|<1/2,$$
where $\uu{\kappa}^s=\frac 1 2 (\uu{\kappa} + \uu{\kappa}^T)$, then there exists a unique stationary solution $\psi_\infty$ which satisfies: $\forall \XX \in {\cal B}(0,\sqrt{b})$,
\begin{equation}\label{eq:estim_psi}
\left| \nabla \left (
    \ln\left( \frac{\psi_\infty(\XX)}{\exp(-\Pi(\XX))}\right)  \right) -
    2 \uu{\kappa}^s \XX \right|
    \leq \frac{2  \sqrt{b} \,
| [\uu{\kappa},\uu{\kappa}^T] |}{1 - 2|\uu{\kappa}^s|},
\end{equation}
where $[.,.]$ is the commutation bracket:
    $[\uu{\kappa},\uu{\kappa}^T]=\uu{\kappa}\uu{\kappa}^T -
    \uu{\kappa}^T\uu{\kappa}$.
Using this result, one may prove that $\U$ converges exponentially fast to $\U_\infty$ in the
$L^2_{\xx}$-norm, and the entropy $\dps{\int_{{\mathcal D}} \int_{{\mathcal B}(0, \sqrt{b})} 
  \ln\left( \frac{\psi}{\psi_\infty}\right) \psi }$ converges exponentially fast to 0, under an additional assumption on $\uu{\kappa}$:
 $$M^2 b^2 \exp(4 b M) + C_{\rm PI}({\mathcal D}) |\uu{\kappa}^s| < 1$$
where $C_{\rm PI}({\mathcal D})$ is the constant in the Poincaré inequality~\eqref{eq:PI} and 
$M=2|\uu{\kappa}^s|  + \frac{2 \,
| [\uu{\kappa},\uu{\kappa}^T] |}{1 - 2|\uu{\kappa}^s|}$.

\subsubsection{Entropy estimates for macroscopic models}

In this section, we are interested in macroscopic models, such as the Oldroyd-B model, or the FENE-P model. The aim is to analyze the longtime behaviour of such models, using ideas developed in the similar study for micro-macro models. In the following, we assume that $\U=0 \text{ on } \partial {\mathcal D}$,
but some generalizations in the spirit of the latter section are possible.

We have seen in Section~\ref{sec:estim_ent} that the analysis of the longtime behaviour of micro-macro models relies on the {\em free energy} $\frac{\Rey}{2} \int_{\cal D} |\U|^2  + \frac{\epsilon}{\We}  \int_{\cal D} \int_{\R^d} \ln \left( \frac{\psi}{\psi_\infty} \right)  \psi$ rather than the {\em energy} $\frac{\Rey}{2} \int_{\cal D} |\U|^2  + \frac{\epsilon}{\We}  \int_{\cal D} \int_{\R^d} \Pi  \, \psi$. Since we mentioned that some macroscopic models (such as the Oldroyd-B model for example) are equivalent to some micro-macro model, it is thus natural to ask how the results obtained for micro-macro models translate to macroscopic models. We refer to~\cite{leonov-92,beris-edwards-94,oettinger-05,wapperom-hulsen-98,hu-lelievre-07} for more details.

We first consider the Oldroyd-B model~\eqref{eq:OB}, which is equivalent to the Hookean dumbbell model. To simplify the presentation, let us rewrite the model in terms of the so-called conformation tensor
\begin{equation}
\uu{A} = \frac{\We}{\epsilon} \ttau + \I
\end{equation}
which writes in the micro-macro formulation as the positive symmetric tensor: $\uu{A} = \E (\XX_t \otimes \XX_t)$. The Oldroyd-B model rewrites in terms of~$(u,p,\uu{A})$:
\begin{equation}\label{eq:macmac_sig}
\left\{
\begin{aligned}
&\Rey \left(\frac{\partial \U}{\partial t} + \U \cdot \nabla \U\right) =  -\nabla p + (1-\epsilon) \Delta \U +  \frac{\epsilon}{\We}\div(\uu{A}), \\
&\div (\U) = 0, \\
&\frac{\partial \uu{A}}{\partial t} + \U \cdot \nabla \uu{A} = \nabla \U \uu{A} + \uu{A} \nabla \U^T -\frac{1}{\We}(\uu{A} -\I).
\end{aligned}
\right.
\end{equation}
To mimic the derivation of~\eqref{eq:estim_NRJ}, we, on the one hand, multiply  by $\U$ and integrate over ${\mathcal D}$  the equation on $\U$  in~\eqref{eq:macmac_sig} (which yields~\eqref{eq:Ec}):
\begin{equation}\label{eq:Ec_sigma}
\frac{\Rey}{2} \frac{d}{dt} \int_{\mathcal D} |\U|^2 + (1 - \epsilon) \int_{\mathcal D} |\nabla \U|^2 = - \frac{\epsilon}{\We} \int_{\mathcal D} \uu{A} : \nabla \U
\end{equation}
and, on the other hand, take the trace of the equation on $\uu{A}$ and integrate it over~${\mathcal D}$:
\begin{equation}\label{eq:Ep_sigma}
 \frac{d}{dt} \int_{\mathcal D} \tr\uu{A} = 2 \int_{\mathcal D} \uu{A} : \nabla \U - \frac{1}{\We} \int_{\mathcal D} \tr\uu{A} + \frac{1}{\We}d |{\mathcal D}|,
\end{equation}
where $d$ denotes the dimension and $|{\mathcal D}|$ the Lebesgue measure of ${\mathcal D}$. By linear combination of~\eqref{eq:Ec_sigma} and~\eqref{eq:Ep_sigma}, we obtain:
\begin{equation}\label{eq:estim_NRJ_OB}
\frac{d}{dt}\left( \frac{\Rey}{2} \int_{{\cal D}} |\U|^2 +
    \frac{\epsilon}{2 \We} \int_{{\cal D}}
    \tr \uu{A} \right) + (1-\epsilon) \int_{{\cal D}} |\nabla \U|^2  +\frac{\epsilon}{2
    \We^2} \int_{\cal D} \tr \uu{A} =  \frac{ \epsilon}{2\We^2} d |{\mathcal D}|.
\end{equation}
This equality is equivalent to~\eqref{eq:estim_NRJ} for Hookean dumbbells
($\Pi(\XX)=|\XX|^2 / 2$). It thus shares the same drawbacks as~\eqref{eq:estim_NRJ} for the study of the longtime behaviour, related to the positive right-hand-side in~\eqref{eq:estim_NRJ_OB}.

To write a free energy estimate of the type~\eqref{eq:estim_ent}, one first establishes the additional identity:
\begin{equation} \label{eq:ent_sigma}
\frac{d}{dt}\int_{{\cal D}} \tr \ln \uu{A} = \frac{1}{\We} \int_{{\cal D}} \tr(\uu{A}^{-1}-\I).
\end{equation}
This identity is obtained from the equation on $\uu{A}$ in~\eqref{eq:macmac_sig}, by contraction with $\uu{A}^{-1}$ and integration over ${\mathcal D}$, using the Jacobi identity:
\begin{equation}\label{eq:jacobi}
\left( \left( \frac{\partial}{\partial t} + \U \cdot \nabla \right) \uu{A} \right):\uu{A}^{-1}= \left( \frac{\partial}{\partial t} + \U \cdot \nabla \right) ( \tr \ln \uu{A} ).
\end{equation}
Combining \eqref{eq:Ec_sigma}, \eqref{eq:Ep_sigma} and~\eqref{eq:ent_sigma}, we obtain:
\begin{equation}\label{eq:estim_ent_OB}
\begin{aligned}
\frac{d}{dt}& \left( \frac\Rey2 \int_{\mathcal D} |\U|^2 +
  \frac{\epsilon}{2\We}\int_{\mathcal D}\tr(\uu{A}-\ln\uu{A} - \I) \right) \\
&+(1-\epsilon)\int_{\mathcal D}|\nabla \U|^2 + \frac{\epsilon}{2\We^2}\int_{\mathcal D}\tr(\uu{A} + \uu{A}^{-1} - 2\I)=0.
\end{aligned}
\end{equation}
Equation~\eqref{eq:estim_ent_OB} is equivalent to~\eqref{eq:estim_ent}. In particular, using that, for any symmetric positive matrix $\uu{M}$,
$$0 \le \tr(\uu{M} -\ln \uu{M} - \I) \le \tr(\uu{M} + \uu{M}^{-1} - 2\I),$$
and using the Poincaré inequality~\eqref{eq:PI}, one obtains the exponential convergence of $(\U, \uu{A})$ to the equilibrium states $(\U_\infty,\uu{A}_\infty)=(0,\I)$.

\medskip
This result on the Oldroyd-B model may be employed as a guideline to derive suitable energy estimates useful for the study of the longtime behaviour of other macroscopic models, even if they are not known to be equivalent to a micro-macro model of the form~\eqref{eq:micmac}. For the FENE-P model~\eqref{eq:FENE-P} for example, one obtains:
\begin{equation}\label{eq:estime_FENE-P}
\begin{aligned}
\frac{d}{dt}\left( \frac{\Rey}{2} \int_{{\cal D}} |\U|^2 +
    \frac{\epsilon}{2\We} \int_{{\cal D}}
   \left(- \ln(\det \uu{A}) - b \ln \left( 1 - \tr(\uu{A})/b\right) \right) \right) + (1-\epsilon) \int_{{\cal D}} |\nabla \U|^2  \\
 +\frac{\epsilon}{2
    \We^2} \int_{\cal D}\left(\frac{ \tr (\uu{A}) }{(1 -
    \tr(\uu{A})/b)^2} - \frac{2d}{1 -  \tr(\uu{A})/b} +
  \tr(\uu{A}^{-1})\right)   = 0.
\end{aligned}
\end{equation}
Using next the Poincaré inequality~\eqref{eq:PI} and that for any symmetric positive matrix $\uu{M}$ (with size $d \times d$), 
\begin{align*}
0 \le - \ln(\det(\uu{M})) &- b \ln \left( 1 - \tr(\uu{M})/b\right) + (b+d) \ln
\left( \frac{b}{b+d} \right) \nonumber \\
&\leq \left(\frac{ \tr (\uu{M}) }{(1 -
    \tr(\uu{M})/b)^2} - \frac{2d}{1 -  \tr(\uu{M})/b} +
  \tr(\uu{M}^{-1})\right),
\end{align*}
one obtains that the free energy $$\frac{\Rey}{2} \int_{{\cal D}} |\U|^2 +
    \frac{\epsilon}{2\We} \int_{{\cal D}}
   \left(- \ln(\det \uu{A}) - b \ln \left( 1 - \tr(\uu{A})/b\right)
   \right)$$ decreases exponentially fast to~$0$.

Besides the study of the longtime behaviour, the interest of such estimates is twofold. They can be used to propose new stability criteria for numerical schemes (see the notion of free energy dissipative schemes introduced in Section~\ref{sec:FE_scheme} below). These estimates can also be used to establish new existence results (see Section~\ref{sec:exist_pde_micmac}).

\subsection{Existence results}

We would like to now present some results concerning existence and uniqueness of models for viscoelastic fluids. Notice beforehand that systems like~\eqref{eq:macmac_sig} modelling purely macroscopically non-Newtonian fluids typically include the Navier-Stokes equations, with the additional
term $\div \uu{A}$ in the right-hand side. The equation on $\uu{A}$ is essentially a
transport equation and, formally, $\uu{A}$ has at best the regularity
of $\nabla \U$. The term $\div \uu{A}$ in the right-hand side
of the momentum equation is thus unlikely to bring more regularity on $\U$. It is
then expected that the study of these coupled systems contains at least
the well-known difficulties of the Navier-Stokes equations. Recall that
for the three-dimensional Navier-Stokes equations, it is known that
global-in-time weak solutions exist but the regularity, and thus the
uniqueness, of such solutions for appropriate data is only known locally in time. We will see below that basically, the picture is similar for some models for viscoelastic fluids.

As mentioned in Section~\ref{sec:math_gen}, the difficulties in the mathematical analysis of such models are related to the {\em transport terms} (in
addition to $\U \cdot \nabla \U$, we now have $\U \cdot \nabla
  \XX_t$ and $\U \cdot \nabla \psi$), the  {\em nonlinear terms} either coming from
  the coupling between variables, or inherently contained in
  the equations defining the stress (due to the non-linear entropic force $\FF$ in micro-macro models, for example).

\subsubsection{Existence results for partial differential equation formulations: macroscopic models}

The state-of-the-art of the mathematical knowledge for the Oldroyd-B model~\eqref{eq:macmac_sig} is existence of local-in-time strong solutions. Example of results for such macroscopic models are contributions by  M.~Renardy
in~\cite{renardy-90} (local-in-time existence for an abstract model that covers the specific Oldroyd-B case),  
by  C.~Guillopé and J.C.~Saut~\cite{guillope-saut-90-a,guillope-saut-90-b} (existence
results for less regular solutions for non-zero viscosity of the solvent), by  E.~Fernandez-Cara, F.~Guillen and
R.R.~Ortega~\cite{fernandez-cara-guillen-ortega-02} and references therein (local-in-time well-posedness in Sobolev spaces), by  F.-H.~Lin, C.~Liu and P.W.~Zhang~\cite{lin-liu-zhang-05} (local-in-time existence and uniqueness results and global-in-time existence and uniqueness results for small data).

The first global-in-time existence result is due to P.-L.~Lions and
N.~Masmoudi~\cite{lions-masmoudi-00}, where an Oldroyd-like model is considered, with the corotational derivative~\eqref{eq:corot_der} rather than the upper convected derivative  on the stress tensor (see Remark~\ref{rem:UCM} above). The proof relies on the fact that for the corotational derivative, some $L^p$ estimates can be obtained on the stress, independently of $\U$.

In a recent work~\cite{masmoudi-10-b}, N.~Masmoudi shows global-in-time existence of weak solutions to the FENE-P model~\eqref{eq:FENE-P}, and other nonlinear macroscopic models (Giesekus model and Phan-Thien-Tanner model). The proof is based on the entropy estimate~\eqref{eq:estime_FENE-P}, and a very fine analysis of how compactness propagates in the system.

\subsubsection{Existence results for partial differential equation formulations: micro-macro models}\label{sec:exist_pde_micmac}

We now turn to the micro-macro models in their Fokker-Planck formulation~\eqref{eq:micmac-FP}. The  analysis of such micro-macro models for
polymeric fluids has begun with an early work by M.~Renardy~\cite{renardy-91}.

The difficulties present for the purely macroscopic models and discussed
above are also present {\it mutatis mutandis} in the
multiscale models.

Several existence and uniqueness results for the coupled system
involving the Fokker-Planck equation have been obtained. For some local existence and uniqueness results, we refer to M.~Renardy~\cite{renardy-91}, T.~Li, H.~Zhang and
P.W.~Zhang~\cite{li-zhang-zhang-04}, H.~Zhang and
P.W.~Zhang~\cite{zhang-zhang-06},  N.~Masmoudi~\cite{masmoudi-08}. The latter author has also shown a global in time existence result for initial data close to equilibrium
(see also F.-H.~Lin, C.~Liu and P.W.~Zhang~\cite{lin-liu-zhang-07} for
a prior, more restricted result).

Global existence results have also been obtained for some closely related
problems:
\begin{itemize}
\item {\em Existence results for a regularized version}: In a series of paper~\cite{barrett-schwab-suli-05,barrett-suli-07,barrett-suli-11}, J.W.~Barrett and E.~S\"uli obtain global
  existence results 
  using space regularized versions. For example, the
  velocity $\U$ in the Fokker Planck equation is replaced by a smoothed
  velocity, and the same smoothing operator is used on the stress tensor
  $\uu{\tau}_p$ in the right-hand side of the momentum equations. Alternatively, a small diffusion term (with respect to the space variable) is introduced in the Fokker-Planck equation. See also
  L.~Zhang, H.~Zhang and P.W.~Zhang~\cite{zhang-zhang-zhang-08}.
\item {\em Existence results with a corotational derivative}:
  In J.W.~Barrett, C.~Schwab and
  E.~S\"uli~\cite{barrett-schwab-suli-05,barrett-suli-07} (again with some
  regularizations) and P.-L.~Lions
  and N.~Masmoudi~\cite{lions-masmoudi-07,masmoudi-08} (without any
  regularizations), the authors obtain global-in-time existence results
  replacing $\nabla \U$ in the Fokker-Planck equation by $\frac{\nabla
    \U - \nabla \U^T}{2}$. More precisely,
  in~\cite{lions-masmoudi-07}, a global-in-time existence result of weak
  solutions is obtained in dimension 2 and 3, while in~\cite{masmoudi-08}, it is proved that
  in dimension 2, there exists a unique global-in-time strong
  solution. A related recent result by F.-H.~Lin, P.~Zhang and Z.~Zhang is~\cite{lin-zhang-zhang-08}.
\end{itemize}

The most striking existence result has been obtained by N. Masmoudi recently, who was able to prove the global existence of weak solutions for the FENE model, see~\cite{masmoudi-10-a}.

We would like also to mention the related
works~\cite{constantin-05,constantin-fefferman-titi-zarnescu-07,constantin-masmoudi-08}
by P.~Constantin, C.~Fefferman, N.~Masmoudi and E.S.~Titi, on existence
results for coupled Navier-Stokes Fokker-Planck micro-macro models.

\subsubsection{Existence results for stochastic differential equation formulations}

\begin{figure}[htbp]
\begin{center}
\includegraphics[width=80mm,angle=0]{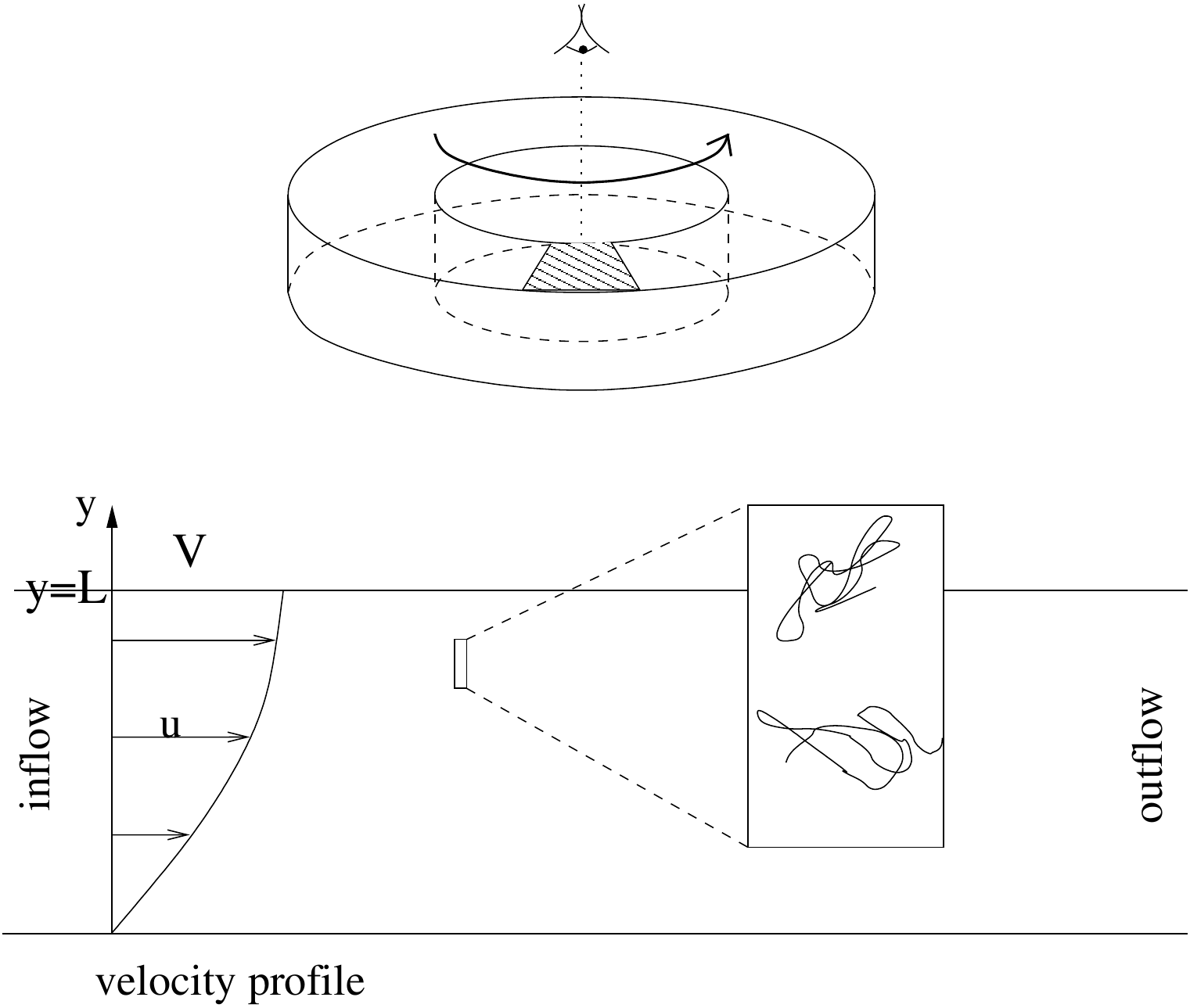}
\end{center}
\caption{
Schematic representation of a rheometer. On an infinitesimal
  angular portion, seen from above, the flow is a simple shear flow
  (Couette flow) confined between two planes with velocity profile
  $(u(t,y),0)$.}\label{fig:couette}
\end{figure}

We now turn to system \eqref{eq:micmac}, which is
quite difficult to study in full generality. For preliminary arguments, a standard simplification is to consider simple shear flows (see in particular in  M.~Laso and H.C.~\"Ottinger~\cite{laso-ottinger-93}, J.C.~Bonvin
and M.~Picasso~\cite{bonvin-picasso-99}, C.~Guillopé and
J.C.~Saut~\cite{guillope-saut-90-b},  W.~E, T.~Li and P.W.~Zhang~\cite{e-li-zhang-02}). In such a case, the velocity is in the $x$-direction, and only depends on the $y$-variable. Such flows are typically studied in rheometers, see Figure~\ref{fig:couette}.

%%%%%%%%%%%%%%%%%%%%%%%%%%%%%%%%%%%%%%%%%%%%%%%%%%%%%%%%%%%%%%%%%%%%%%%%%%
%%%%%%%%%%%%%%%%%%%%%%%%%%%%%%%%%%%%%%%%%%%%%%%%%%%%%%%%%%%%%%%%%%%%%%%%%%

In this simple setting, the system~\eqref{eq:micmac} writes:
\begin{equation}\label{eq:micmac-shear}
\left\lbrace
\begin{array}{l}
\displaystyle {\Rey  \frac{\partial u}{\partial t}(t,y) - (1-\epsilon)
  \frac{\partial^2 u}{\partial y^2}(t,y) = \frac{\partial \tau}{\partial
  y}(t,y),} \\
\displaystyle {\tau(t,y)=\frac{\epsilon}{\We}  \E(P_t(y)
  F_Q(\XX_t(y)))},\\ \noalign{\vskip 6pt}
\displaystyle  d P_t(y)  =  \frac{\partial u}{\partial y} (t,y) Q_t(y) \, dt  - \frac{1}{2 \We}
F_P(\XX_t(y)) \, dt + \frac{1}{\sqrt{\We}} d W^x_t,\\
\displaystyle  d Q_t(y)  =    - \frac{1}{2 \We}
F_Q(\XX_t(y)) \, dt + \frac{1}{\sqrt{\We}} d W^y_t,
\end{array}
\right.
\end{equation}
where $\tau$ is the off-diagonal component of the stress $\ttau$, $(P_t(y),Q_t(y))$ are the two components of the stochastic process
$\XX_t(y)$, and $(F_P,F_Q)$ the two components of the force~$\FF$.

For Hookean dumbbells, the global-in-time existence and uniqueness is easily obtained, since $Q_t$ is known independently of $(u,P_t)$, so that the system becomes linear, see the work by B.~Jourdain, C.~Le~Bris and
T.~Lelièvre~\cite{jourdain-lelievre-le-bris-01}. 

For the FENE force, two new difficulties have to be addressed: first, the stochastic differential equation contains an explosive drift term and second,
even in a shear flow,  the coupling term $\nabla \U \XX_t$ is genuinely
nonlinear. We refer to  B.~Jourdain and
T.~Lelièvre~\cite{jourdain-lelievre-02} and to
B.~Jourdain, C.~Le~Bris and
T.~Lelièvre~\cite{jourdain-lelievre-le-bris-03} for the first studies in
this setting and in particular the proof of a local-in-time existence
and uniqueness result. A surprising result is that if $b$ (the maximum extensibility) is not large enough, the entropic force is not sufficient to prevent the dumbbell to go out of the centered ball with radius $\sqrt{b}$.

For a similar result in a more general setting (3-dimensional flow) and
forces with polynomial growth, we refer to W.~E, T.~Li and P.W.~Zhang~\cite{e-li-zhang-04}. The
authors prove a local-in-time existence and uniqueness result in high
Sobolev spaces. We also refer to A.~Bonito, Ph.~Clément and
M.~Picasso~\cite{bonito-clement-picasso-06-a} for existence results for
Hookean dumbbells, neglecting the advection terms. 

More generally, it is important to notice that, when the velocity field
is not sufficiently regular, and similarly to the situation seen for the
Fokker-Planck setting above, it is
difficult to give a sense to the transport term in the stochastic differential equation (which is
actually a Stochastic {\em Partial} Differential Equation). We refer
to C.~Le~Bris and P.-L.~Lions~\cite{le-bris-lions-04,le-bris-lions-08}
for works in this direction.

\section{Numerical methods and numerical analysis}\label{sec:num}

Most of the numerical methods employed in practice to simulate
models for viscoelastic fluids are based upon a finite element
discretization in space (see however R.~Owens and
T.~Phillips~\cite{owens-phillips-02} for
spectral methods) and  a finite difference discretization in time (usually Euler schemes), with a decoupled
computation of $(\U,p)$ and $\uu{\tau}$. More precisely, at each timestep, the
equation for $(\U,p)$ is first solved, given  the current stress tensor
$\uu{\tau}$. This allows to update the velocity. Next, the equation
for $\uu{\tau}$ (possibly the underlying stochastic model) is solved, and the stress is updated.

\subsection{Generalities}\label{sec:num_gen}

Of course, the difficulties
raised by the discretization of the models are, as always,
reminiscent of the difficulties of the mathematical analysis. 
Here again, as mentioned above, the treatment of the multiscale problem
 necessarily requires a good knowledge of the treatment of the purely
 macroscopic model. 
An overview of the numerical difficulties encountered when simulating
purely macroscopic models for non Newtonian fluids may be found in  R.~Keunings~\cite{keunings-89},
F.P.T.~Baaijens~\cite{baaijens-98}, and R.~Owens and
T.~Phillips~\cite{owens-phillips-02}. 

Let us mention in particular:
\begin{itemize}
\item An inf-sup condition is needed between the discretization space for $\ttau$ and that for $\U$ (in the limit $\epsilon \to 1$). This requires either appropriate discretization spaces (satisfying some inf-sup conditions, see~\cite{marchal-crochet-87}) or stabilization methods (see~\cite{guenette-fortin-99,bonvin-picasso-stenberg-01}).
\item The discretization of the advection terms needs to be adequately performed. Typically, one may use again stabilization methods~\cite{fortin-fortin-89} or a numerical characteristic method~\cite{bonvin-00,wapperom-keunings-legat-00,min-yoo-choi-01}. These difficulties are  prominent for high
Reynolds number (which is not practically relevant in the context of viscoelastic
fluid simulations) or for high Weissenberg number (which \emph{is} relevant).
\item The discretization of the nonlinear term raises difficulties. In particular, treating in the equation on the stress the velocity explicitly and the stress implicitly leads to an ill-posed problem if the Weissenberg number becomes large.
\end{itemize}

We have mentioned that two of these difficulties are prominent for large
Weissenberg number. It indeed appears that numerical methods become
unstable in this latter regime. This is the so-called
High Weissenberg Number Problem\index{High Weissenberg Number Problem}
(HWNP) already mentioned in Section~\ref{sec:mic_vs_mac}. Many works are related to the HWNP (we refer for example
to R.~Owens and
T.~Phillips~\cite[Chapter 7]{owens-phillips-02}). The  HWNP is
certainly not only related to the discretization scheme. It has indeed
been observed that for some geometries, the critical Weissenberg number
(above which the scheme is unstable) decreases with the mesh step size
 (see R.~Keunings~\cite{keunings-00}), which could indicate a loss of
 regularity for the continuous solution itself
 (see D.~Sandri~\cite{sandri-99}). It is still an open problem to precisely
 characterize the HWNP, and to distinguish between instability coming
 from the model itself, or its discretization. For the
 theoretical study of the limit
$\We \to \infty$, we refer to M.~Renardy~\cite[Chapter 6]{renardy-00}.

In the following, we mainly focus on the discretization techniques for micro-macro models. There are two approaches: discretizing the Fokker-Planck version~\eqref{eq:micmac-FP}, or  discretizing the stochastic version~\eqref{eq:micmac}. Let us start with a section on macroscopic models and free energy dissipative schemes.

\subsection{Discretization of macroscopic models: free energy dissipative schemes}\label{sec:FE_scheme}

In this section, we would like to discuss one peculiar aspect of the discretization of macroscopic models, such as the Oldroyd-B model or the FENE-P model. The question we would like to address is the following: what are the necessary ingredients such that a numerical scheme for the Oldroyd-B model~\eqref{eq:macmac_sig} has similar dissipative properties in terms of free energy as the continuous version, see~\eqref{eq:estim_ent_OB}. This would ensure longtime stability of the numerical scheme.

This question has interesting practical consequences since it is often the case that the stationary state is obtained as longtime computation on the time-dependent problem. Besides, as already mentioned above, the discretization of models for viscoelastic fluids is known to be difficult due to some stability problems when the Weissenberg number becomes large, at least in some specific geometries. These unstable behaviours may have two origins: the model itself, or the discretization method. In the first case, the unstable behaviour would be related to some singularities of the continuous solution which appear in finite time. For the Oldroyd-B model for example, it seems that the model itself is sometimes questionable, see~\cite{rallison-hinch-88,thomases-shelley-07,bajaj-pasquali-prakash-08}. This may be seen as a counterpart of the fact that the Hookean dumbbell can extend to infinity, a highly unphysical fact. Other cases are unclear. A numerical scheme recently proposed in~\cite{fattal-kupferman-05,hulsen-fattal-kupferman-05} and based on a reformulation of~\eqref{eq:macmac_sig} in terms of the logarithm of the tensor $\uu{A}$ seems to stay stable for Weissenberg numbers larger than the usual threshold observed for standard numerical schemes. In this context, one aim of the study of the free energy dissipative structure of the numerical scheme is to help to distinguish between stability problems due to the numerical scheme, and stability problems originated form the model itself.

Let us introduce the standard variational formulation used for a finite element discretization of the Oldroyd-B model~\eqref{eq:macmac_sig}: find $(\U,p,\uu{A})$ such that for all $(\V,q,\uu{B})$,
\begin{equation}\label{eq:FV-OB}
\begin{aligned}
\int_{\mathcal D} \Bigg( \Rey \left(\frac{\partial \U}{\partial t} + \U\cdot \nabla \U\right)\cdot \V +
(1-\epsilon) \nabla \U : \nabla \V - p \, \div \V +\frac{\epsilon}{\We}\uu{A} :\nabla \V + q \, \div  \U \\
+ \left(\frac{\partial \uu{A}}{\partial t}+ \U\cdot\nabla \uu{A}\right):\uu{B} - ( \nabla \U \uu{A} +
\uu{A} \nabla \U^T):\uu{B} + \frac{1}{\We}(\uu{A}-\I):\uu{B} \Bigg)= 0.
\end{aligned}
\end{equation}
To obtain the free energy estimate~\eqref{eq:estim_ent_OB}, one in principle just needs to take $(\V,q,\uu{B})=\left(\U,p,\frac{\epsilon}{2 \We}(\I - \uu{A}^{-1})\right)$ test functions. This is obviously not possible for any arbitrary finite element space.

A numerical schemes for a which an estimate  similar to~\eqref{eq:estim_ent_OB} holds at the discrete level is the following: using Scott-Vogelius finite elements ($(\P_d)^2 \times \P_{d-1,\rm disc}$ in dimension $d=2$ or $d=3$) for the velocity-pressure, $\P_0$ finite elements for the tensor $\uu{A}$, and a discretization of the advection terms by the characteristic method, or a discontinuous Galerkin method. As an example, in the latter case, the variational formulation is: find $(\U_h^{n+1},p_h^{n+1},\uu{A}_h^{n+1}) \in (\P_d)^2 \times \P_{d-1,disc} \times (\P_0)^3$ such that for all $(\V,q,\uu{B})\in (\P_d)^2 \times \P_{d-1,disc} \times (\P_0)^3$:
\begin{equation}\label{eq:FV-DG}
\begin{aligned}
 \sum_{k=1}^{N_K} \int_{K_k} \Bigg( \Rey \left( \frac{\U_h^{n+1} - \U_h^n}{\delta t} + \U_h ^n \cdot \nabla \U_h^{n+1} \right) \cdot \V
- p_h^{n+1} \div \V + q \div \U_h^{n+1}
\\
 + (1-\epsilon) \nabla \U_h^{n+1} : \nabla \V +\frac{\epsilon}{\We} \uu{A}_h^{n+1}: \nabla \V + \frac{1}{\We} (\uu{A}_h^{n+1}-\I):\uu{B}
\\
+ \left(  \frac{\uu{A}_h^{n+1}-\uu{A}_h^n}{\delta t} \right):\uu{B}
- \left( \nabla \U_h^{n+1} \uu{A}_h^{n+1} + \uu{A}_h^{n+1}(\nabla \U_h^{n+1})^T \right):\uu{B} \Bigg)
\\
 + \sum_{j=1}^{N_E} \int_{E_j} |\U_h^n \cdot \uu{n}| \jump{\uu{A}_h^{n+1}} : \uu{B}^+=0,
\end{aligned}
\end{equation}
where $(K_k)_{1 \le k \le N_K}$ denote the mesh cells, $(E_j)_{1 \le j \le N_E}$ the internal edges of the mesh, $\uu{n}$ is the normal on the edge, $\uu{B}^+$ the downstream value of $\phi$ (for the vector field $\U_h^n$) and $\jump{\uu{A}_h^{n+1}} = (\uu{A}_h^{n+1})^+ - (\uu{A}_h^{n+1})^-$ is the jump of $\uu{A}_h^{n+1}$ through the edge. The following stability result then holds:
\begin{theo}\label{theo:stab-disc}
For a given mesh, there exists a constant $c_0>0$ such that for all timestep sizes $\delta t < c_0$, there exists a unique solution to~\eqref{eq:FV-DG}, with $\uu{A}_h^{n+1}$ a positive definite matrix. Moreover, this solution satisfies the following free energy estimate:
\begin{equation}
\begin{aligned}
&F_h^{n+1} - F_h^n  + \int_{\mathcal D} \frac{\Rey}{2} |\U_h^{n+1}-\U_h^n|^2 \\
 &+ \delta t \int_{\mathcal D} \left( (1-\epsilon) |\nabla \U_h^{n+1}|^2 + \frac{\epsilon}{2 {\We}^2} \tr \left(\uu{A}_h^{n+1}+(\uu{A}_h^{n+1})^{-1}-2 \I \right)\right) \le 0
\end{aligned}
\end{equation}
where
$$F_h^n=\frac\Rey 2 \int_{\mathcal D} |\U_h^n|^2 +
  \frac{\epsilon}{2\We}\int_{\mathcal D}\tr(\uu{A}_h^n-\ln\uu{A}_h^n - \I).$$
\end{theo}
We refer to~\cite{boyaval-lelievre-mangoubi-09} for more details on free energy dissipative schemes. The main conclusions therein are:
\begin{itemize}
\item[(i)] The discretization method of the advection terms plays a crucial role in the proofs. A numerical characteristic method or a discontinuous Galerkin method are adequate discretization techniques.
\item[(ii)] The discretization space for the stress should contain piecewise constant functions. This is related to the fact that $P_0$ is a stable finite element space under the application $\uu{A} \mapsto \frac{\epsilon}{2 \We}(\I - \uu{A}^{-1})$ (which is the test function we would like to choose to recover the free energy estimate).
\item[(iii)] It is possible to obtain a free energy dissipative scheme without limitation on the timestep size using the logarithmic formulation of the stress, as recently introduced in~\cite{fattal-kupferman-05,hulsen-fattal-kupferman-05} (whereas a CFL condition is required for standard formulations in terms of $\uu{A}$, see Theorem~\ref{theo:stab-disc}). Moreover, any solution to the logarithmic formulation satisfies a free energy estimate (and this is not true for the standard formulation in $\uu{A}$). This is maybe related to the fact that these schemes seem to be more stable than standard ones, but definite conclusions in this direction require further studies.
\end{itemize}

\subsection{Discretization of the Fokker-Planck version: solving high-dimensional partial differential equations}

Spectral methods are typically used to discretize the Fokker Planck equation in~\eqref{eq:micmac-FP} (see A.~Lozinski~\cite{lozinski-03} or J.K.C.~Suen, Y.L.~Joo and R.C.~Armstrong~\cite{suen-joo-armstrong-02}). It is not easy to find
a suitable variational formulation of the Fokker Planck equation, and an
appropriate discretization that satisfy the natural constraints on
the probability density $\psi$ (namely non negativity, and
normalization).  We refer
to C.~Chauvi{\`e}re and A.~Lozinski~\cite{chauviere-lozinski-04,lozinski-chauviere-03} for appropriate
discretizations in the FENE case. One major difficulty in the
discretization of Fokker-Planck equations is when the configurational space is
high-dimensional.

The main difficulty with this approach comes from the high-dimensionality of the function $\psi$. For the dumbbell model in dimension $3$, $\psi$ is a function of $7$ scalar variables. When the polymer chain is modelled by a chain of $N$ beads linked by
springs, the Fokker-Planck equation is a parabolic equation posed on a
$3N$-dimensional domain, and $\psi$ is a function of $4+3N$ scalar variables. Some numerical methods
have been developed to discretize such high dimensional problems. The
idea is to use an appropriate Galerkin basis, whose dimension does not
explode when dimension grows. We refer
to P.~Delaunay, A.~Lozinski and
R.G.~Owens~\cite{delaunay-lozinski-owens-07}, T.~von~Petersdorff and
C.~Schwab~\cite{von-petersdorff-schwab-04}, H.-J.~Bungartz and M.~Griebel~\cite{bungartz-griebel-04}
for the sparse-tensor product
approach, to L.~Machiels, Y.~Maday, and A.T.~Patera~\cite{machiels-maday-patera-01} for the reduced basis approach and
to A.~Ammar, B.~Mokdad, F.~Chinesta and R.~Keunings~\cite{ammar-mokdad-chinesta-keunings-06,ammar-mokdad-chinesta-keunings-07}
for a method coupling a sparse-tensor product discretization with a
reduced approximation basis approach.

We more precisely describe the latter method, which is an interesting nonlinear approximation method. For simplicity, we concentrate on the Poisson equation in dimension $2$, but the generalization to any symmetric problem on a high-dimensional cylindrical domain is straightforward.

Consider the problem:
\begin{equation}\label{eq:lapl}
\text{Find $g \in H^1_0(\Omega) $ such that }\left\{
\begin{array}{rl}
-\Delta g = f &\text{ in $\Omega$},\\
g=0& \text{ on $\partial \Omega$},
\end{array}
\right.
\end{equation}
where  $f \in L^2(\Omega)$, $\Omega=\Omega_x
\times \Omega_y$ with $\Omega_x \subset
\R$ and $\Omega_y \subset \R$ two bounded domains. The bottom line of the algorithm proposed in~\cite{ammar-mokdad-chinesta-keunings-06,ammar-mokdad-chinesta-keunings-07}  is to look for the solution as a sum of tensor products $g = \sum_{k \ge 1} r_k \otimes s_k$ (where $r_k \otimes s_k(x,y)=r_k(x)s_k(y)$) computing iteratively the terms in the sum: set $f_0=f$, and iterate on $n \ge 1$,
\begin{enumerate}
\item Find $r_n \in H^1_0(\Omega_x)$ and $s_n \in H^1_0(\Omega_y)$ such that: for all functions $(r,s) \in H^1_0(\Omega_x)\times H^1_0(\Omega_y)$
\begin{equation}\label{eq:LRA_EL_FV}
\int_{\Omega} \nabla (r_n \otimes s_n) \cdot \nabla (r_n \otimes s + r \otimes
s_n)  = \int_{\Omega} f_{n-1} (r_n \otimes s + r \otimes s_n).
\end{equation}
\item Set $f_{n}=f + \Delta \sum_{k=1}^n r_k \otimes s_k$.
\item If $\|f_{n}\|_{H^{-1}(\Omega)} \ge \varepsilon$, then go to iteration $n+1$, otherwise stop.
\end{enumerate}
Notice that this method can indeed be generalized to high dimensional Poisson problems since the problem~\eqref{eq:LRA_EL_FV} in dimension $N$ consists in solving a (non-linear) system of $N$ equations. This has to be compared with a classical Galerkin approach, which would lead to a linear problem with dimension that exponentially  scales in $N$. This method reportedly gives reliable results in high dimension, see~\cite{ammar-mokdad-chinesta-keunings-06,ammar-mokdad-chinesta-keunings-07} and ~\cite{nouy-07,nouy-le-maitre-09} for applications in the context of uncertainty quantification.

A fundamental remark to understand the algorithm is that the variational problem~\eqref{eq:LRA_EL_FV} is the Euler equation associated to the minimization problem: Find $r_n \in H^1_0(\Omega_x)$ and $s_n \in H^1_0(\Omega_y)$ such that,
\begin{equation}\label{eq:LRA_varo}
(r_n,s_n)=\arg\min_{(r,s) \in H^1_0(\Omega_x) \times H^1_0(\Omega_y)} \left(
  \frac{1}{2}\int_\Omega |\nabla (r \otimes s)|^2 - \int_\Omega f_{n-1}^o  \,(r \otimes s) \right).
\end{equation}
Then, as shown in~\cite{le-bris-lelievre-maday-09}, the proposed algorithm falls within a class of so-called {\em Greedy Algorithms} introduced in nonlinear approximation theory, see~\cite{barron-cohen-dahmen-devore-08,davis-mallat-avellaneda-97,devore-temlyakov-96,temlyakov-08}. Using results from~\cite{devore-temlyakov-96}, it is possible to prove the convergence of the algorithm with rate $O(1/\sqrt{n})$. For convergence results for nonlinear problems (minimization of strongly convex functionals), we refer to~\cite{cances-ehrlacher-lelievre-10}.

The convergence of such algorithms for non-symmetric problems is, to the best of our knowledge, an open question.

\subsection{Discretization of the stochastic version}

As mentioned in the previous section, discretizing the Fokker-Planck version~\eqref{eq:micmac-FP} of the micro-macro models is certainly not the most natural approach especially when the microscopic model involves numerous degrees of freedom (a chain of springs rather than a dumbbell to model the polymer chain, for example).

\subsubsection{The CONNFFESSIT method}

The standard approach to discretize~\eqref{eq:micmac} couples a finite element
discretization and a Monte Carlo technique. In this context, this approach is called CONNFFESSIT\index{CONNFFESSIT} for {\em
  Calculation Of Non-Newtonian Flow: Finite Elements and Stochastic
  SImulation Technique} (see M.~Laso and H.C.~\"Ottinger~\cite{laso-ottinger-93}).

To discretize the expectation
in~\eqref{eq:micmac}, a Monte Carlo method is employed: at each
macroscopic point $\xx$ ({\em i.e.} at each node of the mesh once the
problem is discretized) many replicas (or realizations)  $(\XX^{k,K}_t)_{1 \le k \le K}$ of
the stochastic process $\XX_t$ are simulated, driven by independent
Brownian motions $(\WW^k_t)_{k \ge 1}$, and the stress tensor is
obtained as an empirical mean over these processes:
\begin{equation}\label{eq:tau_MC}
\uu{\tau}^K=\frac{\epsilon}{\We}\left( \frac{1}{K} \sum_{k=1}^K
  \XX^{k,K}_t \otimes \FF(\XX^{k,K}_t) - \I \right).
\end{equation}
This approximate stress is then inserted in the mass and momentum conservation laws, to update the velocity and the pressure.

One important feature
of such a discretization is that, at the discrete level, all
the unknowns $(\U,p,\uu{\tau})$ become {\em random variables}. The consequence is that variance is typically the
bottleneck for the accuracy of the method. In particular, variance
reduction methods are relevant and important.

\subsubsection{Convergence of the CONNFFESSIT method}

For simplicity, we concentrate on a simple shear flow~\eqref{eq:micmac-shear}. In this case, and for Hookean dumbbells, the CONNFFESSIT method writes as follows: at time $t_n=n \delta t$ ($\delta t$ being the fixed timestep size), for a given velocity field $\ou_h^n \in V_h$ and dumbbell configurations $(\oP^{k,n}_h,\oQ^{k,n}_h)_{1 \le k \le K}\in \partial_yV_h$, compute $\ou_h^{n+1} \in V_h$ and $(\oP^{k,n+1}_h,\oQ^{k,n+1}_h)_{1 \le k \le K}\in \partial_yV_h$ solution to:
$$ \left\lbrace
 \begin{aligned} 
 &\frac{\Rey}{\delta t} \int \left( \ou_h^{n+1} - \ou_h^n \right) v_h +
 (1-\epsilon)\int \partial_{y} \ou_h^{n+1} \partial_{y} v_h = -\int \otau_h^n
 \partial_y v_h, \qquad  \forall v_h \in V_h, \\
 &\otau_h^n = \frac{\epsilon}{\We} \frac 1 K \sum_{k=1}^K \left(\oP^{k,n}_h \oQ_h^{k,n}\right),\\
 &\oP^{k,n+1}_h= \oP^{k,n}_h+\left( - \frac{1}{2 \We} \oP^{k,n}_h + \partial_y \ou_h^{n+1}
 \oQ_h^{k,n} \right) \, \delta t + \frac{1}{\sqrt{ \We}} \left(W^{x,k}_{t_{n+1}} -W^{x,k}_{t_{n}}\right),\\
 &\oQ_h^{k,n+1}= \oQ_h^{k,n} + \left(-\frac{1}{2 \We} \oQ_h^{k,n} \right) \, \delta t
  + \frac{1}{\sqrt{ \We}}\left(W^{y,k}_{t_{n+1}} -W^{y,k}_{t_{n}}\right).
 \end{aligned}
 \right.
$$
Here $(W^{x,k}_t,W^{y,k}_t)_{1 \le k \le K}$ is a collection of independent two-dimensional Brownian motions, $V_h$ is a finite element space (think of $P1$ finite elements, namely continuous piecewise affine functions) and $\partial_y V_h$ denotes the space of derivatives with respect to $y$ of functions in $V_h$ (think of $P0$ functions, namely piecewise constant functions). For such a problem, a typical result of numerical
analysis first proved in B.~Jourdain, C.~Le~Bris and
T.~Lelièvre~\cite{jourdain-lelievre-le-bris-01} (see also W.~E, T.~Li and
P.W.~Zhang~\cite{e-li-zhang-02}) is the following error estimate:
 \begin{equation*}
 \bigg|\bigg|u(t_n)-
     \ou_h^n \bigg|\bigg|_{L^2_y(L^2_\omega)} \!\!\!\!\!
 +
 \bigg|\bigg|\E(P_{t_n}Q_{t_n}) - \frac{1}{K} \sum_{k=1}^K \oP^{k,n}_h \oQ^{k,n}_h  \bigg|\bigg|_{L^1_y(L^1_\omega)}
 \leq 
 C \left(\Delta y +\delta t +\frac{1}{\sqrt{K}}\right),
 \end{equation*}
where $\Delta y$ is the size of the mesh associated to the finite element space $V_h$. The precise result actually requires a cut-off procedure, which is applied with very small probability in the limit $\Delta t \to 0$ or $K \to \infty$, see~\cite{jourdain-lelievre-le-bris-01} for the details.

The main difficulties for the proof originate 
from the following facts:
\begin{itemize}
\item The velocity $\ou_h^n$ is a random variable. The energy estimate at the continuous
level cannot be directly translated into an energy estimate at the discrete
level due to some correlation between random variables. The stability of the scheme is thus not trivial to obtain, and actually requires an almost sure
  bound on $\oQ_h^{k,n}$ which is obtained through a cut-off procedure.
\item The end-to-end vectors $(\oP_h^{k,n}, \oQ_h^{k,n})_{1 \le k \le K}$
  are {\em coupled} random variables (even though the driving Brownian motions
  $(W^{x,k}_t,W^{y,k}_t)_{1 \le k \le K}$ are independent).
\end{itemize}
 For an extension of these
results to a more general geometry and a discretization by a finite
difference scheme, we refer to T.~Li and
P.W.~Zhang~\cite{li-zhang-06}. A convergence result in space and
time may be found in  A.~Bonito, Ph.~Clément and
M.~Picasso~\cite{bonito-clement-picasso-06-b}.

\subsubsection{Variance reduction for the CONNFFESSIT method}\label{sec:VarRed}

We would like to now discuss two variance reduction methods for the CONNFFESSIT discretization.

\paragraph{The control variate method}

The first method is the control variate method. Let us start by a general presentation of the method.

Assume that, in full generality, one wants to compute the expectation $\E(Z)$ of a random variable $Z$ using the Monte Carlo method. The idea of the control variate method is to write 
$$\E(Z) = \E(Z- \alpha Y) + \alpha \E(Y)$$
(where $Y$, the so-called {\em control variate}, is a random variable with a {\em known} average, and $\alpha$ is a deterministic parameter to be fixed) and to approximate the expectation $\E(Z- \alpha Y)$ using a Monte Carlo method. 
For the approach to be efficient, we need the control variate $Y$ to be such that
$$\Var (Z- \alpha Y) \ll \Var (Z).$$
The optimal value of $\alpha$ can be explicitly computed. It is  $\alpha^*=\frac{{\rm Covar}(Z,Y)}{\Var(Y)}$ so that
$\Var (Z- \alpha^* Y) = \Var(Z) ( 1 - \rho (Z,Y)^2)$  with  $$\rho(Z,Y)=\frac{{\rm Covar}(Z,Y)}{\sqrt{\Var(Y) \Var(Z)}} \in [-1,1].$$
This shows that the optimal case is $|\rho(Z,Y)|=1$ ({\em i.e.}, up to an additive constant $ Y = Z - \E(Z)$) but this requires to know $\E(Z)$ which is precisely what we want to compute. The worst case is $\rho(Z,Y)=0$ (that is $Y$ and $Z$ are uncorrelated) in which case the control variate does not help in reducing the variance. The practical question is then how to build a control variate $Y$ sufficiently correlated to $Z$, but such that $\E(Y)$ is explicitly known (or sufficiently easy to compute) ?

\medskip

In the context of micro-macro models, the Monte Carlo method is used to approximate the stress $\ttau$ (see~\eqref{eq:tau_MC}) and one can build control variates using two companion systems: (i) the same problem at equilibrium or (ii) the same problem with a force for which a macroscopic equivalent is known (Hookean force, or the FENE-P closure approximation, for example). For the FENE model for example, one may write:
\begin{align*}
\E\left( \frac{\XX_t \otimes
      \XX_t}{1- | \XX_t |^2 / b} \right) &=\E\left( \frac{\XX_t \otimes
      \XX_t}{1- | \XX_t |^2 / b} - \tXX_t \otimes \tilde{\FF}(\tXX_t) \right)\\
& \quad +  \E\left( \tXX_t \otimes \tilde{\FF}(\tXX_t)\right),
\end{align*}
with either:
\begin{itemize}
\item (control variate at equilibrium) $\tilde{\FF}(\XX)=\FF(\XX)=\frac{\XX}{1-|\XX|^2/b}$ and $d\tXX_t + \U \cdot \nabla \tXX_t \, dt = -\frac{1}{2 \We} \tilde{\FF}(\tXX_t) \, dt + \frac{1}{\sqrt{\We}} d\WW_t$.
\item (Hookean dumbbell as control variate) $\tilde{\FF}(\tXX)=\tXX$ and $d\tXX_t+ \U \cdot \nabla \tXX_t \, dt = \left(\nabla \U \tXX_t -\frac{1}{2 \We} \tilde{\FF}(\tXX_t) \right) \, dt + \frac{1}{\sqrt{\We}} d\WW_t$.
\end{itemize}
In both cases, $ \E\left( \tXX_t \otimes \tilde{\FF}(\tXX_t)\right)$ can be computed using deterministic approaches: it is either the stress at equilibrium (thus an analytically known constant), or the solution to the Oldroyd-B partial differential equation. Of course, for the method to be effective (namely for the control variate to be indeed correlated to the original random variable), the Brownian motion driving $\tXX_t$ needs to be {\em the same} as the Brownian motion driving~$\XX_t$.

\paragraph{Dependency of the Brownian motion upon the space variable}

Another variance reduction technique is very specific to micro-macro models such as~\eqref{eq:micmac}. It consists in using the space correlation of the driving Brownian motion as a way to control the variance of the result.

We again consider Hookean dumbbells in a shear flow (see~\eqref{eq:micmac-shear}):
\begin{equation}\label{eq:micmac-shear-HD}
\left\lbrace
\begin{array}{l}
\displaystyle {\Rey  \frac{\partial u}{\partial t}(t,y) - (1-\epsilon)
  \frac{\partial^2 u}{\partial y^2}(t,y) = \frac{\partial \tau}{\partial
  y}(t,y),} \\
\displaystyle {\tau(t,y)=\frac{\epsilon}{\We}  \E(P_t(y) Q_t)},\\ \noalign{\vskip 6pt}
\displaystyle  d P_t(y)  =  \frac{\partial u}{\partial y} (t,y) Q_t \, dt  - \frac{1}{2 \We} P_t(y) \, dt + \frac{1}{\sqrt{\We}} d W^x_t,\\
\displaystyle  d Q_t  =    - \frac{1}{2 \We} Q_t \, dt + \frac{1}{\sqrt{\We}} d W^y_t,
\end{array}
\right.
\end{equation}
One important remark (valid in any geometry and for any force) is that the macroscopic variables $(\U,p,\ttau)$ do not depend on the correlation in space of the Brownian motions. This owes to the fact that $\ttau$ is defined as a pointwise average. At the discrete level, however, the variance of the results does depend on this correlation, which thus may be seen as a adjustable numerical parameter that may help reducing the variance of the results.

We study this problem in~\cite{jourdain-le-bris-lelievre-04} on the simple system~\eqref{eq:micmac-shear-HD}. In particular we justify theoretically the numerical observations~\cite{halin-lielens-keunings-legat-98,bonvin-picasso-99}: using on different mesh elements mutually independent Brownian motions increases the variance on the velocity and decreases the variance on the stress compared to using a Brownian motion constant in space. Loosely speaking, the fact the variance on the velocity is large with mutually independent Brownian motions is related to the fact that a derivative with respect to space of the Brownian motion appears through the term $\frac{\partial \tau}{\partial y}$ in the right hand side of the equation on $u$. In some formal sense, this is a derivative of a white noise. The fact that the variance on the stress is smaller with mutually independent Brownian motions is related to the fact that the stress appears to be an integral in space of the Brownian motions, and a sum of independent random variables as a smaller variance than a sum of the same random variable (namely $\Var\left(\sum_{i=1}^I G^i \right) < \Var\left(\sum_{i=1}^I G^1
\right)$  where $G^i$ are i.i.d.). The main conclusions of~\cite{jourdain-le-bris-lelievre-04} are:
 \begin{itemize}
 \item
  The variance of the results comes from  an interplay
 between the space discretized operators and the covariance of the
 Brownian motion in space.
 \item The minimum of the variance of $u$ is obtained for a Brownian
   constant in space.
 \item The minimum of the variance of $\tau$ is {\em not} obtained with 
    mutually independent Brownian motions. One can further reduce
   the variance using a Brownian motion $W_t$ multiplied alternatively
   by $+1$ or $-1$ from one cell to another (in the spirit of variance reduction through antithetic variates).
 \end{itemize}

\subsubsection{A stochastic reduced basis approach for micro-macro problems}

We conclude this article summarizing a recent work~\cite{boyaval-lelievre-10} on a {\em reduced basis approach} for Monte Carlo computations. The motivation is the following: in the CONNFFESSIT method, the expected values:
$$\E(\XX_t \otimes \FF(\XX_t))$$
where $\XX_t$ satisfies the stochastic differential equation:
$$d\XX_t + \U \cdot \nabla_{\uu{x}} \XX_t \, dt = \left( \nabla \U \cdot \XX_t -
  \frac{1}{2 {\We}} \FF(\XX_t) \right) dt +\frac{1}{\sqrt{{\We}}} d\WW_t$$
have to be computed through a Monte Carlo procedure {\em for many values of the velocity gradient}. The principle is to use this ``many query'' context to build a variance reduction method.

The two building blocks are: the control variate method already explained above in Section~\ref{sec:VarRed} and the reduced basis method, which we now outline.

\paragraph{The reduced basis method} The reduced basis method~\cite{prudhomme-rovas-veroy-maday-patera-turinici-02,boyaval-le-bris-lelievre-maday-nguyen-patera-10} has been proposed to efficiently solve partial differential equations, typically of the form $-\div(A(\lambda) \nabla u) = f$, when the solution $u(\lambda)$ has to be repeatedly computed for many values of the parameter $\lambda \in \Lambda$.

The method consists: 
\begin{itemize}
\item in an {\em offline stage}, in building a reduced basis $$X_N={\rm span}(u(\lambda_1), \ldots, u(\lambda_N))$$ for appropriately chosen values of the parameter $\lambda$, and
\item in an {\em online stage}, in looking for a variational solution to the original problem in $X_N$, using a Galerkin method.
\end{itemize}
In the offline stage, the parameter values $\lambda_i$ are computed using a greedy algorithm: choose a finite set $\Lambda_{\rm trial} \subset \Lambda$, and $\lambda_1 \in \Lambda_{\rm trial}$. Then, for $n \ge 0$, set
$$\lambda_{n+1} \in \arg\sup_{\lambda \in \Lambda_{\rm trial}} \Delta_n(\lambda)$$
where $\Delta_n(\lambda)$ is an {\em a posteriori} estimator of the error made on the output, when approximating $u(\lambda)$ by $u_n(\lambda) \in {\rm span}(u(\lambda_1), \ldots, u(\lambda_n))$.

We purposely omit to mention other aspects, and in particular the fact that, to be efficient, the method requires some orthogonalization procedure of the reduced basis. This approach has proven to be useful in many contexts. A satisfactory accuracy is typically achieved for $N$ of the order of $10$. In summary, the main tools used in the reduced basis methodology are:
\begin{itemize}
\item {\em A two-stage offline-online strategy} (which is efficient either in a many-query context, or for real-time computing);
\item the idea to use solutions at given values of the parameter to build {\em a reduced basis} on which solutions for other values are expanded;
\item A procedure to {\em select the best linear combination} on a given reduced basis;
\item An {\em a posteriori estimator} used online and offline to evaluate the error.
\item {\em A greedy algorithm} to select offline the best values of the parameter among a trial sample to build the reduced basis;
\end{itemize}
We now describe a method, based upon the same ideas, to build a control variate to reduce the variance for Monte Carlo estimations by empirical means of parametrized random variables.

\paragraph{A reduced basis method for Monte Carlo computations}

Consider the problem of approximating, for many values of $\lambda$, $\E(Z^\lambda)$ by $$E_M=\frac{1}{M} \sum_{i=1}^M (Z^\lambda_i - Y^\lambda_i),$$ where $(Z^\lambda_i,Y^\lambda_i)$ are i.i.d. We use the empirical variance $$V_M=\frac{1}{M} \sum_{i=1}^M (Z^\lambda_i - Y^\lambda_i - E_M)^2$$
as an {\em a posteriori} estimator of the error.
As mentioned above, an optimal, unfortunately inaccessible, control variate is $Y^\lambda=Z^\lambda- \E(Z^\lambda)$.

We thus propose the following two-stage algorithm. In the offline stage, optimal control variates are computed, using a large number of samples:
$$Y^{\lambda_n} = Z^{\lambda_n} -\frac{1}{M_{\rm large}} \sum_{i=1}^{M_{\rm large}} Z^{\lambda_n}_i.$$
In the online stage, we use
$$\tilde{Y}^\lambda=\sum_{n=1}^N \alpha^*_n Y^{\lambda_n}$$
where
$(\alpha^*_n)=
\arg\inf_{(\alpha_n)} \Var\left( Z^\lambda - \sum_{n=1}^N \alpha_n Y^{\lambda_n} \right)$,  as a control variate.
The optimal $(\alpha^*_n)$ are solution to a least square problem. Then, the expected values are approximated online using empirical means over a small number of samples (typically $M_{\rm large}=100 M_{\rm small}$). Finally, the parameters $\lambda_n$ in the offline stage are computed using a greedy algorithm, in order to minimize the variance.

\begin{figure}[htbp]
\centerline{\includegraphics[trim = 10mm 0mm 20mm 10mm,clip,scale=.3]{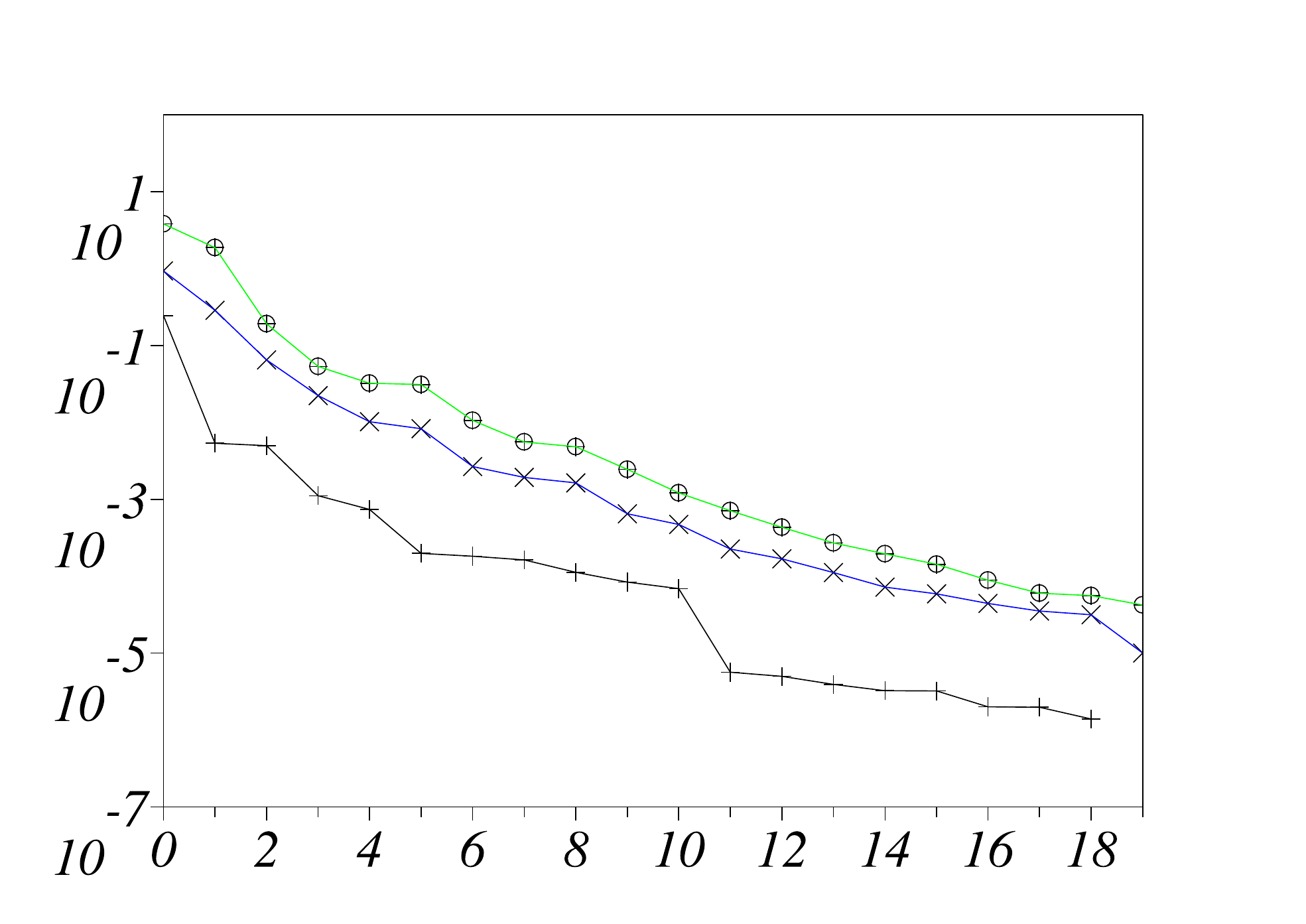}
\includegraphics[trim = 10mm 0mm 20mm 10mm,clip,scale=.3]{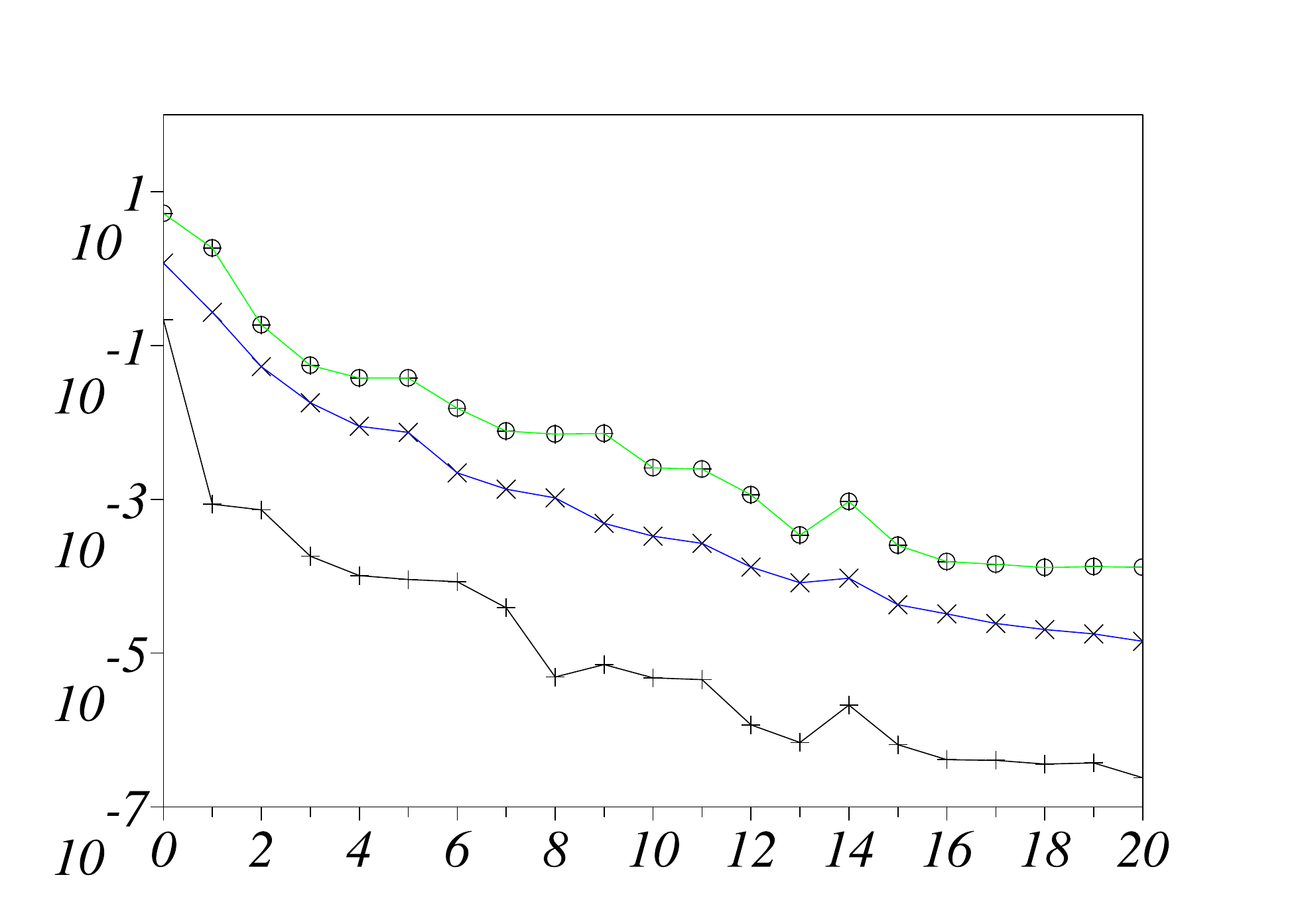}}
\caption{Maximum, median and minimum of the variance over the samples, in the offline stage, over $\Lambda_{\rm trial}$ (left) and in the online stage, over $\Lambda$ (right), as a function of the size of the reduced basis.}\label{fig:RB}
\end{figure}

In~\cite{boyaval-lelievre-10}, we apply this technique to the FENE model, $\E(Z^\lambda)$ being the stress tensor and the parameters $\lambda$  being the components of the gradient of the velocity field. One representative numerical result is displayed on Figure~\ref{fig:RB}. As seen on this Figure, the conclusions are:
\begin{itemize}
\item Using the control variate method with a $20$-dimensional reduced basis of control variates, the variance is approximately divided by a factor of $10^4$ in the mean for large test samples of parameters, at least in the applications we experimented. As a consequence, the reduced-basis approaches allows to approximately divide the online computation time by a factor of $10^2$, while maintaining 
the confidence intervals for the output expectation at the same value than without reduced basis.
\item The greedy algorithm seems to efficiently select the values of the parameter $\lambda$ since in the online stage, on a large set $\Lambda$ of values for the parameter $\lambda$, the reduction on the variance is basically of the same order as the one observed in the offline stage, on the trial sample $\Lambda_{\rm trial}$.
\end{itemize}
There exists another version of the algorithm specialized to the case when $Z^\lambda$ is a function of some solution to a parametrized stochastic differential equation. We refer to~\cite{boyaval-lelievre-10} for more details.

% \bibliography{biblio_HD,biblio_MD,ma_biblio}

\begin{thebibliography}{100}

\bibitem{ammar-mokdad-chinesta-keunings-06}
A.~Ammar, B.~Mokdad, F.~Chinesta, and R.~Keunings.
\newblock A new family of solvers for some classes of multidimensional partial
  differential equations encountered in kinetic theory modeling of complex
  fluids.
\newblock {\em J. Non-Newtonian Fluid Mech.}, 139:153--176, 2006.

\bibitem{ammar-mokdad-chinesta-keunings-07}
A.~Ammar, B.~Mokdad, F.~Chinesta, and R.~Keunings.
\newblock A new family of solvers for some classes of multidimensional partial
  differential equations encountered in kinetic theory modeling of complex,
  part {II}: Transient simulation using space-time separated representations.
\newblock {\em J. Non-Newtonian Fluid Mech.}, 144:98--121, 2007.

\bibitem{ABC-00}
C.~An\'e, S.~Blach\`ere, D.~Chafa\"i, P.~Foug\`eres, I.~Gentil, F.~Malrieu,
  C.~Roberto, and G.~Scheffer.
\newblock {\em Sur les in\'{e}galit\'{e}s de Sobolev logarithmiques}.
\newblock Soci\'{e}t\'{e} Math\'{e}matique de France, 2000.
\newblock In French.

\bibitem{arnold-markowich-toscani-unterreiter-01}
A.~Arnold, P.~Markowich, G.~Toscani, and A.~Unterreiter.
\newblock On convex {S}obolev inequalities and the rate of convergence to
  equilibrium for {F}okker-{P}lanck type equations.
\newblock {\em Comm. Part. Diff. Eq.}, 26:43--100, 2001.

\bibitem{baaijens-98}
F.P.T. Baaijens.
\newblock Mixed finite element methods for viscoelastic flow analysis: a
  review.
\newblock {\em J. Non-Newtonian Fluid Mech.}, 79:361--385, 1998.

\bibitem{bajaj-pasquali-prakash-08}
M.~Bajaj, M.~Pasquali, and J.R. Prakash.
\newblock Coil-stretch transition and the breakdown of computations for
  viscoelastic fluid flow around a confined cylinder.
\newblock {\em J. Rheol.}, 52:197--223, 2008.

\bibitem{barrett-schwab-suli-05}
J.W. Barrett, C.~Schwab, and E.~S\"{u}li.
\newblock Existence of global weak solutions for some polymeric flow models.
\newblock {\em Math. Models and Methods in Applied Sciences}, 15(6):939--983,
  2005.

\bibitem{barrett-suli-07}
J.W. Barrett and E.~S\"{u}li.
\newblock Existence of global weak solutions to kinetic models for dilute
  polymers.
\newblock {\em Multiscale Model. Simul.}, 6(2):506--546, 2007.

\bibitem{barrett-suli-11}
J.W. Barrett and E.~S\"{u}li.
\newblock Existence and equilibration of global weak solutions to finitely
  extensible nonlinear bead-spring chain models for dilute polymers.
\newblock {\em Math. Models and Methods in Applied Sciences}, 2011.
\newblock To appear.

\bibitem{barron-cohen-dahmen-devore-08}
A.R. Barron, A.~Cohen, W.~Dahmen, and R.A. DeVore.
\newblock Approximation and learning by greedy algorithms.
\newblock {\em Annals of Statistics}, 36(1):64--94, 2008.

\bibitem{beris-edwards-94}
A.N. Beris and B.J. Edwards.
\newblock {\em Thermodynamics of flowing systems with internal microstructure}.
\newblock Oxford University Press, 1994.

\bibitem{BCAH-87-1}
R.B. Bird, R.C. Armstrong, and O.~Hassager.
\newblock {\em Dynamics of polymeric liquids}, volume~1.
\newblock Wiley Interscience, 1987.

\bibitem{BCAH-87-2}
R.B. Bird, C.F. Curtiss, R.C. Armstrong, and O.~Hassager.
\newblock {\em Dynamics of polymeric liquids}, volume~2.
\newblock Wiley Interscience, 1987.

\bibitem{bird-dotson-johnson-80}
R.B. Bird, P.J. Dotson, and N.L. Johnson.
\newblock Polymer solution rheology based on a finitely extensible bead-spring
  chain model.
\newblock {\em J. Non-Newtonian Fluid Mech.}, 7:213--235, 1980.
\newblock Errata: {\it J. Non-Newtonian Fluid Mech.}, 8:193 (1981).

\bibitem{blanc-legoll-le-bris-lelievre-10}
X.~Blanc, F.~Legoll, C.~Le~Bris, and T~Leli\`evre.
\newblock Beyond multiscale and multiphysics: Multimaths for model coupling.
\newblock {\em Netw. Heterog. Media}, 5(3):423--460, 2010.

\bibitem{bonito-clement-picasso-06-b}
A.~Bonito, Ph. Cl\'{e}ment, and M.~Picasso.
\newblock Finite element analysis of a simplified stochastic {H}ookean
  dumbbells model arising from viscoelastic flows.
\newblock {\em M2AN Math. Model. Numer. Anal.}, 40(4):785--814, 2006.

\bibitem{bonito-clement-picasso-06-a}
A.~Bonito, Ph. Cl\'{e}ment, and M.~Picasso.
\newblock Mathematical analysis of a stochastic simplified {H}ookean dumbbells
  model arising from viscoelastic flow.
\newblock {\em J. Evol. Equ.}, 6(3):381--398, 2006.

\bibitem{bonvin-picasso-99}
J.~Bonvin and M.~Picasso.
\newblock Variance reduction methods for {CONNFFESSIT}-like simulations.
\newblock {\em J. Non-Newtonian Fluid Mech.}, 84:191--215, 1999.

\bibitem{bonvin-picasso-stenberg-01}
J.~Bonvin, M.~Picasso, and R.~Stenberg.
\newblock {GLS} and {EVSS} methods for a three fields {S}tokes problem arising
  from viscoelastic flows.
\newblock {\em Comp. Meth. Appl. Mech. Eng.}, 190:3893--3914, 2001.

\bibitem{bonvin-00}
J.C. Bonvin.
\newblock {\em Numerical simulation of viscoelastic fluids with mesoscopic
  models}.
\newblock PhD thesis, Ecole Polytechnique F\'ed\'erale de Lausanne, 2000.
\newblock Available at {\tt http://library.epfl.ch/theses/?nr=2249}.

\bibitem{boyaval-le-bris-lelievre-maday-nguyen-patera-10}
S.~Boyaval, C.~Le~Bris, T~Leli\`evre, Y.~Maday, N.C. Nguyen, and A.T. Patera.
\newblock Reduced basis techniques for stochastic problems.
\newblock {\em Arch. Comput. Method. E.}, 2010.
\newblock to appear.

\bibitem{boyaval-lelievre-10}
S.~Boyaval and T.~Leli\`evre.
\newblock A variance reduction method for parametrized stochastic differential
  equations using the reduced basis paradigm.
\newblock {\em Commun. Math. Sci.}, 8(3):735--762, 2010.

\bibitem{boyaval-lelievre-mangoubi-09}
S.~Boyaval, T.~Leli{\`e}vre, and C.~Mangoubi.
\newblock Free-energy-dissipative schemes for the {O}ldroyd-{B} model.
\newblock {\em ESAIM-Math. Model. Num.}, 43:523--561, 2009.

\bibitem{braack-ern-03}
M.~Braack and A.~Ern.
\newblock A posteriori control of modeling errors and discretization errors.
\newblock {\em Multiscale Model. Simul.}, 1(2):221--238, 2003.

\bibitem{bungartz-griebel-04}
H.-J. Bungartz and M.~Griebel.
\newblock Sparse grids.
\newblock {\em Acta Numer.}, 13:147--269, 2004.

\bibitem{cances-ehrlacher-lelievre-10}
E.~Canc\`es, V.~Ehrlacher, and T.~Leli\`evre.
\newblock Convergence of a greedy algorithm for high-dimensional convex
  nonlinear problems, 2010.
\newblock Available at {\tt http://arxiv.org/abs/1004.0095}.

\bibitem{chauviere-lozinski-04}
C.~Chauvi{\`e}re and A.~Lozinski.
\newblock Simulation of dilute polymer solutions using a {Fokker-Planck}
  equation.
\newblock {\em Computers and fluids}, 33(5-6):687--696, 2004.

\bibitem{constantin-05}
P.~Constantin.
\newblock Nonlinear {F}okker-{P}lanck {N}avier-{S}tokes systems.
\newblock {\em Commun. Math. Sci.}, 3(4):531--544, 2005.

\bibitem{constantin-fefferman-titi-zarnescu-07}
P.~Constantin, C.~Fefferman, A.~Titi, and A.~Zarnescu.
\newblock Regularity of coupled two-dimensional nonlinear {F}okker-{P}lanck and
  {N}avier-{S}tokes systems.
\newblock {\em Commun. Math. Phys.}, 270(3):789--811, 2007.

\bibitem{constantin-masmoudi-08}
P.~Constantin and N.~Masmoudi.
\newblock Global well posedness for a {S}moluchowski equation coupled with
  {N}avier-{S}tokes equations in 2d.
\newblock {\em Commun. Math. Phys.}, 278:179--191, 2008.

\bibitem{davis-mallat-avellaneda-97}
G.~Davis, S.~Mallat, and M.~Avellaneda.
\newblock Adaptive greedy approximations.
\newblock {\em Constr. Approx.}, 13(1):57--98, 1997.

\bibitem{delaunay-lozinski-owens-07}
P.~Delaunay, A.~Lozinski, and R.G. Owens.
\newblock Sparse tensor-product {F}okker-{P}lanck-based methods for nonlinear
  bead-spring chain models of dilute polymer solutions.
\newblock {\em CRM Proceedings and Lecture Notes, Volume 41}, 2007.

\bibitem{devore-temlyakov-96}
R.A. DeVore and V.N. Temlyakov.
\newblock Some remarks on greedy algorithms.
\newblock {\em Adv. Comput. Math.}, 5:173--187, 1996.

\bibitem{doi-edwards-88}
M.~Doi and S.F. Edwards.
\newblock {\em The Theory of Polymer Dynamics}.
\newblock International Series of Monographs on Physics. Oxford University
  Press, 1988.

\bibitem{e-li-zhang-02}
W.~E, T.~Li, and P.W. Zhang.
\newblock Convergence of a stochastic method for the modeling of polymeric
  fluids.
\newblock {\em Acta Mathematicae Applicatae Sinica, English Series},
  18(4):529--536, 2002.

\bibitem{e-li-zhang-04}
W.~E, T.~Li, and P.W. Zhang.
\newblock Well-posedness for the dumbbell model of polymeric fluids.
\newblock {\em Commun. Math. Phys.}, 248:409--427, 2004.

\bibitem{ern-lelievre-07}
A.~Ern and T.~Leli\`evre.
\newblock Adaptive models for polymeric fluid flow simulation.
\newblock {\em C. R. Acad. Sci. Paris, Ser. I}, 344(7):473--476, 2007.

\bibitem{fattal-kupferman-05}
R.~Fattal and R.~Kupferman.
\newblock Time-dependent simulation of viscoelastic flows at high {W}eissenberg
  number using the log-conformation representation.
\newblock {\em J. Non-Newtonian Fluid Mech.}, 126:23--37, 2005.

\bibitem{fernandez-cara-guillen-ortega-02}
E.~Fern{\'a}ndez-Cara, F.~Guill{\'e}n, and R.R. Ortega.
\newblock Mathematical modeling and analysis of viscoelastic fluids of the
  {O}ldroyd kind.
\newblock In P.G. Ciarlet~et al., editor, {\em Handbook of numerical analysis.
  Vol. 8: Solution of equations in $\R^n$ (Part 4). Techniques of scientific
  computing (Part 4). Numerical methods of fluids (Part 2).}, pages 543--661.
  Elsevier, 2002.

\bibitem{fortin-fortin-89}
M.~Fortin and A.~Fortin.
\newblock A new approach for the {FEM} simulation of viscoelastic flows.
\newblock {\em J. Non-Newtonian Fluid Mech.}, 32:295--310, 1989.

\bibitem{grunewald-otto-villani-westdickenberg-09}
N.~Grunewald, F.~Otto, C.~Villani, and M.G. Westdickenberg.
\newblock A two-scale approach to logarithmic {S}obolev inequalities and the
  hydrodynamic limit.
\newblock {\em Ann. Inst. H. Poincaré Probab. Statist.}, 45(2):302--351, 2009.

\bibitem{guenette-fortin-99}
R.~Gu\'enette and M.~Fortin.
\newblock A new mixed finite element method for computing viscoelastic flows.
\newblock {\em J. Non-Newtonian Fluid Mech.}, 60:27--52, 1999.

\bibitem{guillope-saut-90-a}
C.~Guillop\'e and J.C. Saut.
\newblock Existence results for the flow of viscoelastic fluids with a
  differential constitutive law.
\newblock {\em Nonlinear Analysis, Theory, Methods \& Appl.}, 15(9):849--869,
  1990.

\bibitem{guillope-saut-90-b}
C.~Guillop\'e and J.C. Saut.
\newblock Global existence and one-dimensional nonlinear stability of shearing
  motions of viscoelastic fluids of {Oldroyd} type.
\newblock {\em RAIRO Math. Model. Num. Anal.}, 24(3):369--401, 1990.

\bibitem{halin-lielens-keunings-legat-98}
P.~Halin, G.~Lielens, R.~Keunings, and V.~Legat.
\newblock The {Lagrangian} particle method for macroscopic and micro-macro
  viscoelastic flow computations.
\newblock {\em J. Non-Newtonian Fluid Mech.}, 79:387--403, 1998.

\bibitem{hebraud-lequeux-98}
P.~H\'ebraud and F.~Lequeux.
\newblock Mode-coupling theory for the pasty rheology of soft glassy materials.
\newblock {\em Phys. Rev. Lett.}, 81:2934--2937, 1998.

\bibitem{hu-lelievre-07}
D.~Hu and T.~Leli\`evre.
\newblock New entropy estimates for the {O}ldroyd-{B} model, and related
  models.
\newblock {\em Commun. Math. Sci.}, 5(4):906--916, 2007.

\bibitem{hulsen-fattal-kupferman-05}
M.A. Hulsen, R.~Fattal, and R.~Kupferman.
\newblock Flow of viscoelastic fluids past a cylinder at high {W}eissenberg
  number: stabilized simulations using matrix logarithms.
\newblock {\em Journal of Non-Newtonian Fluid Mechanics}, 127(1):27--39, 2005.

\bibitem{jourdain-le-bris-lelievre-04}
B.~Jourdain, C.~Le~Bris, and T.~Leli\`evre.
\newblock On a variance reduction technique for micro-macro simulations of
  polymeric fluids.
\newblock {\em J. Non-Newtonian Fluid Mech.}, 122:91--106, 2004.

\bibitem{jourdain-le-bris-lelievre-otto-06}
B.~Jourdain, C.~Le~Bris, T.~Leli\`evre, and F.~Otto.
\newblock Long-time asymptotics of a multiscale model for polymeric fluid
  flows.
\newblock {\em Archive for Rational Mechanics and Analysis}, 181(1):97--148,
  2006.

\bibitem{jourdain-lelievre-02}
B.~Jourdain and T.~Leli\`evre.
\newblock Mathematical analysis of a stochastic differential equation arising
  in the micro-macro modelling of polymeric fluids.
\newblock In I.M. Davies, N.~Jacob, A.~Truman, O.~Hassan, K.~Morgan, and N.P.
  Weatherill, editors, {\em Probabilistic Methods in Fluids Proceedings of the
  Swansea 2002 Workshop}, pages 205--223. World Scientific, 2003.

\bibitem{jourdain-lelievre-le-bris-01}
B.~Jourdain, T.~Leli\`evre, and C.~Le~Bris.
\newblock Numerical analysis of micro-macro simulations of polymeric fluid
  flows: a simple case.
\newblock {\em Math. Models and Methods in Applied Sciences}, 12(9):1205--1243,
  2002.

\bibitem{jourdain-lelievre-le-bris-03}
B.~Jourdain, T.~Leli\`evre, and C.~Le~Bris.
\newblock Existence of solution for a micro-macro model of polymeric fluid: the
  {FENE} model.
\newblock {\em J. Funct. Anal.}, 209:162--193, 2004.

\bibitem{keunings-89}
R.~Keunings.
\newblock Simulation of viscoelastic fluid flow.
\newblock In C.~Tucker, editor, {\em Fundamentals of Computer Modeling for
  Polymer Processing}, pages 402--470. Hanse, 1989.

\bibitem{keunings-00}
R.~Keunings.
\newblock A survey of computational rheology.
\newblock In D.M.~Binding et~al., editor, {\em Proc. 13th Int. Congr. on
  Rheology}, pages 7--14. British Society of Rheology, 2000.

\bibitem{laso-ottinger-93}
M.~Laso and H.C. \"Ottinger.
\newblock Calculation of viscoelastic flow using molecular models : The
  {CONNFFESSIT} approach.
\newblock {\em J. Non-Newtonian Fluid Mech.}, 47:1--20, 1993.

\bibitem{le-bris-lelievre-09}
C.~{Le Bris} and T.~Leli\`evre.
\newblock {\em Multiscale modelling of complex fluids: A mathematical
  initiation}, volume~66 of {\em Lecture Notes in Computational Science and
  Engineering}.
\newblock Springer, 2009.

\bibitem{le-bris-lelievre-maday-09}
C.~Le~Bris, T.~Leli\`evre, and Y.~Maday.
\newblock Results and questions on a nonlinear approximation approach for
  solving high-dimensional partial differential equations.
\newblock {\em Constructive Approximation}, 30(3):621--651, 2009.

\bibitem{le-bris-lions-04}
C.~Le~Bris and P.L. Lions.
\newblock Renormalized solutions to some transport equations with partially
  {$W^{1,1}$} velocities and applications.
\newblock {\em Annali di Matematica pura ed applicata}, 183:97--130, 2004.

\bibitem{le-bris-lions-08}
C.~Le~Bris and P.L. Lions.
\newblock Existence and uniqueness of solutions to {F}okker-{P}lanck type
  equations with irregular coefficients.
\newblock {\em Comm. Part. Diff. Eq.}, 33(7):1272--1317, 2008.

\bibitem{lee-xu-06}
Y.~Lee and J.~Xu.
\newblock New formulations positivity preserving discretizations and stability
  analysis for non-{N}ewtonian flow models.
\newblock {\em Comput. Methods Appl. Mech. Engrg.}, 195:1180--1206, 2006.

\bibitem{legoll-lelievre-10}
F.~Legoll and T.~Leli{\`e}vre.
\newblock Effective dynamics using conditional expectations.
\newblock {\em Nonlinearity}, 23:2131--2163, 2010.

\bibitem{lelievre-09}
T.~Leli{\`e}vre.
\newblock A general two-scale criteria for logarithmic {S}obolev inequalities.
\newblock {\em J. Funct. Anal.}, 256(7):2211--2221, 2009.

\bibitem{lelievre-rousset-stoltz-book-10}
T.~Leli\`evre, M.~Rousset, and G.~Stoltz.
\newblock {\em Free energy computations: A mathematical perspective}.
\newblock Imperial College Press, 2010.

\bibitem{leonov-92}
A.I. Leonov.
\newblock Analysis of simple constitutive equations for viscoelastic liquids.
\newblock {\em J. non-newton. fluid mech.}, 42(3):323--350, 1992.

\bibitem{li-zhang-zhang-04}
T.~Li, H.~Zhang, and P.W. Zhang.
\newblock Local existence for the dumbbell model of polymeric fluids.
\newblock {\em Comm. Part. Diff. Eq.}, 29(5-6):903--923, 2004.

\bibitem{li-zhang-06}
T.~Li and P.W. Zhang.
\newblock Convergence analysis of {BCF} method for {H}ookean dumbbell model
  with finite difference scheme.
\newblock {\em SIAM MMS}, 5(1):205--234, 2006.

\bibitem{lin-liu-zhang-05}
F.-H. Lin, C.~Liu, and P.W. Zhang.
\newblock On hydrodynamics of viscoelastic fluids.
\newblock {\em Comm. Pure Appl. Math.}, 58(11):1437--1471, 2005.

\bibitem{lin-liu-zhang-07}
F.-H. Lin, C.~Liu, and P.W. Zhang.
\newblock On a micro-macro model for polymeric fluids near equilibrium.
\newblock {\em Comm. Pure Appl. Math.}, 60(6):838--866, 2007.

\bibitem{lin-zhang-zhang-08}
F.-H. Lin, P.~Zhang, and Z.~Zhang.
\newblock On the global existence of smooth solution to the {2-D FENE} dumbell
  model.
\newblock {\em Comm. Math. Phys.}, 277:531--553, 2008.

\bibitem{lions-masmoudi-00}
P.L. Lions and N.~Masmoudi.
\newblock Global solutions for some {O}ldroyd models of non-{N}ewtonian flows.
\newblock {\em Chin. Ann. Math., Ser. B}, 21(2):131--146, 2000.

\bibitem{lions-masmoudi-07}
P.L. Lions and N.~Masmoudi.
\newblock Global existence of weak solutions to micro-macro models.
\newblock {\em C. R. Math. Acad. Sci.}, 345(1):15--20, 2007.

\bibitem{lozinski-03}
A.~Lozinski.
\newblock {\em Spectral methods for kinetic theory models of viscoelastic
  fluids}.
\newblock PhD thesis, Ecole Polytechnique F\'{e}d\'{e}rale de Lausanne, 2003.
\newblock Available at {\tt http://library.epfl.ch/theses/?nr=2860}.

\bibitem{lozinski-chauviere-03}
A.~Lozinski and C.~Chauvi{\`e}re.
\newblock A fast solver for {Fokker-Planck} equation applied to viscoelastic
  flows calculations.
\newblock {\em J. Comp. Phys.}, 189(2):607--625, 2003.

\bibitem{machiels-maday-patera-01}
L.~Machiels, Y.~Maday, and A.T. Patera.
\newblock Output bounds for reduced-order approximations of elliptic partial
  differential equations.
\newblock {\em Comput. Methods Appl. Mech. Engrg.}, 190(26-27):3413--3426,
  2001.

\bibitem{marchal-crochet-87}
J.M. Marchal and M.J. Crochet.
\newblock A new mixed finite element for calculating viscoelastic flows.
\newblock {\em J. Non-Newtonian Fluid Mech.}, 26:77--114, 1987.

\bibitem{masmoudi-10-b}
N.~Masmoudi.
\newblock Global existence of weak solutions to macroscopic models of polymeric
  flows.
\newblock Available at {\tt
  http://www.math.nyu.edu/faculty/masmoudi/fenep.pdf}.

\bibitem{masmoudi-10-a}
N.~Masmoudi.
\newblock Global existence of weak solutions to the fene dumbbell model of
  polymeric flows.
\newblock Available at {\tt http://arxiv.org/abs/1004.4015}.

\bibitem{masmoudi-08}
N.~Masmoudi.
\newblock Well posedness for the {FENE} dumbbell model of polymeric flows.
\newblock {\em Comm. Pure Appl. Math.}, 61(12):1685--1714, 2008.

\bibitem{min-yoo-choi-01}
T.~Min, J.Y. Yoo, and H.~Choi.
\newblock Effect of spatial discretization schemes on numerical solutions of
  viscoelastic fluid flows.
\newblock {\em J. Non-Newtonian Fluid Mech.}, 100:27--47, 2001.

\bibitem{nouy-07}
A.~Nouy.
\newblock A generalized spectral decomposition technique to solve a class of
  linear stochastic partial differential equations.
\newblock {\em Comput. Methods Appl. Mech. Engrg.}, 196:4521--4537, 2007.

\bibitem{nouy-le-maitre-09}
A.~Nouy and O.P. Le~Ma\^itre.
\newblock Generalized spectral decomposition method for stochastic non linear
  problems.
\newblock {\em Journal of Computational Physics}, 228:202--235, 2009.

\bibitem{oden-prudhomme-02}
J.T. Oden and S.~Prudhomme.
\newblock Estimation of modeling error in computational mechanics.
\newblock {\em J. Comput. Phys.}, 182:496--515, 2002.

\bibitem{oden-vemaganti-00}
J.T. Oden and K.S. Vemaganti.
\newblock Estimation of local modeling error and goal-oriented adaptive
  modeling of heterogeneous materials. i. error estimates and adaptive
  algorithms.
\newblock {\em J. Comput. Phys.}, 164:22--47, 2000.

\bibitem{oettinger-95}
H.C. \"Ottinger.
\newblock {\em Stochastic Processes in Polymeric Fluids}.
\newblock Springer, 1995.

\bibitem{oettinger-05}
H.C. \"Ottinger.
\newblock {\em Beyond Equilibrium Thermodynamics}.
\newblock Wiley, 2005.

\bibitem{otto-reznikoff-07}
F.~Otto and M.G. Reznikoff.
\newblock A new criterion for the logarithmic {S}obolev inequality and two
  applications.
\newblock {\em J. Funct. Anal.}, 243:121--157, 2007.

\bibitem{owens-06}
R.G. Owens.
\newblock A new microstructure-based constitutive model for human blood.
\newblock {\em J. Non-Newtonian Fluid Mech.}, 140:57--70, 2006.

\bibitem{owens-phillips-02}
R.G. Owens and T.N. Phillips.
\newblock {\em Computational rheology}.
\newblock Imperial College Press / World Scientific, 2002.

\bibitem{peterlin-66}
A.~Peterlin.
\newblock Hydrodynamics of macromolecules in a velocity field with longitudinal
  gradient.
\newblock {\em J. Polym. Sci. B}, 4:287--291, 1966.

\bibitem{prudhomme-rovas-veroy-maday-patera-turinici-02}
C.~Prud'homme, D.~Rovas, K.~Veroy, Y.~Maday, A.T. Patera, and G.~Turinici.
\newblock Reliable real-time solution of parametrized partial differential
  equations: Reduced-basis output bounds methods.
\newblock {\em Journal of Fluids Engineering}, 124(1):70--80, 2002.

\bibitem{rallison-hinch-88}
J.M. Rallison and E.J. Hinch.
\newblock Do we understand the physics in the constitutive equation ?
\newblock {\em J. Non-Newtonian Fluid Mech.}, 29:37--55, 1988.

\bibitem{renardy-90}
M.~Renardy.
\newblock Local existence of solutions of the {Dirichlet} initial-boundary
  value problem for incompressible hypoelastic materials.
\newblock {\em SIAM J. Math. Anal.}, 21(6):1369--1385, 1990.

\bibitem{renardy-91}
M.~Renardy.
\newblock An existence theorem for model equations resulting from kinetic
  theories of polymer solutions.
\newblock {\em SIAM J. Math. Anal.}, 22:313--327, 1991.

\bibitem{renardy-00}
M.~Renardy.
\newblock {\em Mathematical analysis of viscoelastic flows}.
\newblock SIAM, 2000.

\bibitem{sandri-99}
D.~Sandri.
\newblock Non integrable extra stress tensor solution for a flow in a bounded
  domain of an {Oldroyd} fluid.
\newblock {\em Acta Mech.}, 135(1--2):95--99, 1999.

\bibitem{suen-joo-armstrong-02}
J.K.C. Suen, Y.L. Joo, and R.C. Armstrong.
\newblock Molecular orientation effects in viscoelasticity.
\newblock {\em Annu. Rev. Fluid Mech.}, 34:417--444, 2002.

\bibitem{temlyakov-08}
V.N. Temlyakov.
\newblock Greedy approximation.
\newblock {\em Acta Numerica}, 17:235--409, 2008.

\bibitem{thomases-shelley-07}
B.~Thomases and M.~Shelley.
\newblock Emergence of singular structures in {Oldroyd-B} fluids.
\newblock {\em Phys. Fluids}, 19:103103, 2007.

\bibitem{von-petersdorff-schwab-04}
T.~von Petersdorff and C.~Schwab.
\newblock Numerical solution of parabolic equations in high dimensions.
\newblock {\em M2AN Math. Model. Numer. Anal.}, 38(1):93--127, 2004.

\bibitem{wapperom-hulsen-98}
P.~Wapperom and M.A. Hulsen.
\newblock Thermodynamics of viscoelastic fluids: the temperature equation.
\newblock {\em J. Rheol.}, 42(5):999--1019, 1998.

\bibitem{wapperom-keunings-legat-00}
P.~Wapperom, R.~Keunings, and V.~Legat.
\newblock The backward-tracking lagrangian particle method for transient
  viscoelastic flows.
\newblock {\em J. Non-Newtonian Fluid Mech.}, 91:273--295, 2000.

\bibitem{zhang-zhang-06}
H.~Zhang and P.W. Zhang.
\newblock Local existence for the {FENE}-dumbbell model of polymeric fluids.
\newblock {\em Archive for Rational Mechanics and Analysis}, 2:373--400, 2006.

\bibitem{zhang-zhang-zhang-08}
L.~Zhang, H.~Zhang, and P.W. Zhang.
\newblock Global existence of weak solutions to the regularized
  {H}ookean-dumbbell model.
\newblock {\em Commun. Math. Sci.}, 6(1):83--124, 2008.

\end{thebibliography}
% \bibliographystyle{plain}

\end{document}